\documentclass[3p,fleqn]{elsarticle}

\usepackage[english]{babel}
\usepackage[utf8]{inputenc}
\usepackage[T1]{fontenc}

\usepackage{amsthm,amssymb}
\usepackage[fleqn]{mathtools}
\usepackage{booktabs}
\usepackage{array}
\usepackage{graphicx}
\usepackage{tabu}
\usepackage{caption}
\usepackage{hyperref}
\usepackage[ruled,vlined]{algorithm2e}
\usepackage[nameinlink,capitalize]{cleveref}

\graphicspath{{figures/}}

\hypersetup{
    colorlinks = true
}

\newdefinition{col}{Corollary}

\begin{document}

\begin{frontmatter}
	\title{Information processing constraints in travel behaviour modelling: A generative learning approach}
	
	\author[1]{Melvin Wong\corref{cor1}}
    \ead{melvin.wong@ryerson.ca}
    
    \author[1]{Bilal Farooq}
    \ead{bilal.farooq@ryerson.ca}
    
    \address[1]{Laboratory of Innovations in Transportation, Civil Engineering Department, Ryerson University, 350 Victoria Street, Toronto, Ontario M5B 2K3, Canada}
    \cortext[cor1]{Corresponding author}

    \begin{abstract}
        Travel decisions tend to exhibit sensitivity to uncertainty and information processing constraints.
        These behavioural conditions can be characterized by a generative learning process.
        We propose a data-driven generative model version of rational inattention theory to emulate these behavioural representations.
        We outline the methodology of the generative model and the associated learning process as well as provide an intuitive explanation of how this process captures the value of prior information in the choice utility specification.
        We demonstrate the effects of information heterogeneity on a travel choice, analyze the econometric interpretation, and explore the properties of our generative model.
        Our findings indicate a strong correlation with rational inattention behaviour theory, which suggest that individuals may ignore certain exogenous variables and rely on prior information for evaluating decisions under uncertainty.
        Finally, the principles demonstrated in this study can be formulated as a generalized entropy and utility based multinomial logit model.
    \end{abstract}
    
    \begin{keyword}
    	Information theory \sep generative model \sep rational inattention \sep variational inference
    \end{keyword}
\end{frontmatter}

\section{Introduction}
\label{sec: Introduction}
The classical assumption about modelling travel behaviour data is that individuals have varying unobserved heterogeneity in their choice preferences \citep{mcfaddentrain2000}. 
In recent years, the use of data-driven modelling and integration of behavioural and psychological factors in discrete choice and travel behaviour analysis have become active areas of research \citep{lietal2016,vijkrueger2017,nikolicbierlaire2017}.
In the context of data-driven models, behavioural variations describe the correlation between observed choice attributes and unobserved socio-economic factors using a flexible and tractable model specification. 
These variations include: \textit{decision-protocols}, \textit{choice sets}, \textit{unobserved taste variations} and \textit{unobserved attributes} \citep{gopinath1994}. 
Under these considerations, recent studies on travel behaviour analysis have so far primarily focused on representing heterogeneity in the error correction function and incorporating it into utility based multinomial logit (MNL) models \citep{vijkrueger2017}.
Models such as mixed multinomial logit (MMNL) or latent class (LC) model offers flexibility in representing heterogeneity and substitution patterns.
In addition, recent conceptual frameworks such as the integrated choice and latent variable (ICLV) use individuals' psychometric indicators to represent unobserved behavioural and perception heterogeneity \citep{bolducalvarezdaziano2010}. 
It is also possible to apply a generative machine learning to identify informative latent constructs in travel decision making without subjective behaviour indicators \citep{wong2018modelling,wong2018discriminative}.
However, the true underlying behavioural patterns are often unknown and usually approximated by some pre-determined exogenous indicator variables that would often lead to model misspecification due to lack of complete information, or error in data collection \citep{cherchipolak2005}. 
Furthermore, accurate specification of the underlying distribution assumes individuals have access to all available information regarding the travel activity (e.g. travel times of each mode, knowledge of exact traffic status, etc.).
This information will not always be available to the individual and they might also choose to not consider these variables in their decision making process.
Therefore, statistical variations in the observed data may not exhibit the same underlying properties as with the individuals' behaviour.

A different perspective to explain these heterogeneity manifestations is to consider the element of information processing costs based on rational inattention theory \citep{sims2010,matvejkamckay2015}.
Rational inattention theory is defined as individuals choosing their optimal preference, at the same time considering incomplete information about the choice attributes and relying on their prior beliefs about the choice set.
A typical example would be route choice selection: Individuals tend to ignore most path choices and consider only a few prioritized routes in their choice set \citep{alizadeh2018online}.
These manifestations occur through repeated choice process and prior experiences about the travel routes.
As described in \cite{matvejkamckay2015}, information theoretic approaches do not impose any particular assumptions on what is learned or how they are learned---the structure of the model is estimated through the minimization of decision uncertainty. 
Under this interpretation, a rational inattention model captures the systematic utility and adjusts for prior knowledge and individuals' internal information processing strategy using an entropy term. 
Individuals perceive route choices with heterogeneous prior beliefs and allocate different levels of attention to each alternative.
Consequently, misspecification in classical econometric model estimation can be interpreted as the systematic error between the data observed by the analyst and the true underlying heterogeneous beliefs of the decision makers (which are hidden to the analyst). 

The objective of this research is to model unobserved variations in travel behaviour data by emulating decisions under uncertainty and information processing constraints as a data-driven generative learning process.
We develop a choice model estimation framework with latent constructs that capture information heterogeneity within the data. 
The key difference between our work and previous literature is that we show how rational inattention can be framed as a flexible and extendable generative learning model that emulates the cognitive processes in human behaviour \citep{fosgerauetal2017,fosgeraujiang2019}.
We postulate that realistic behavioural patterns can be modelled using a data-driven generative learning process and we estimate a model to represent the underlying heterogeneity of the data.
Lastly, we provide a quantifiable economic interpretation using latent variables by analyzing the model properties and systematic effects from the latent variable parameters.
This will provide valuable insights into how modern data-driven and deep learning techniques can be exploited to improve travel behaviour modelling.

\begin{figure}[t]
    \centering
    \includegraphics[width=\textwidth]{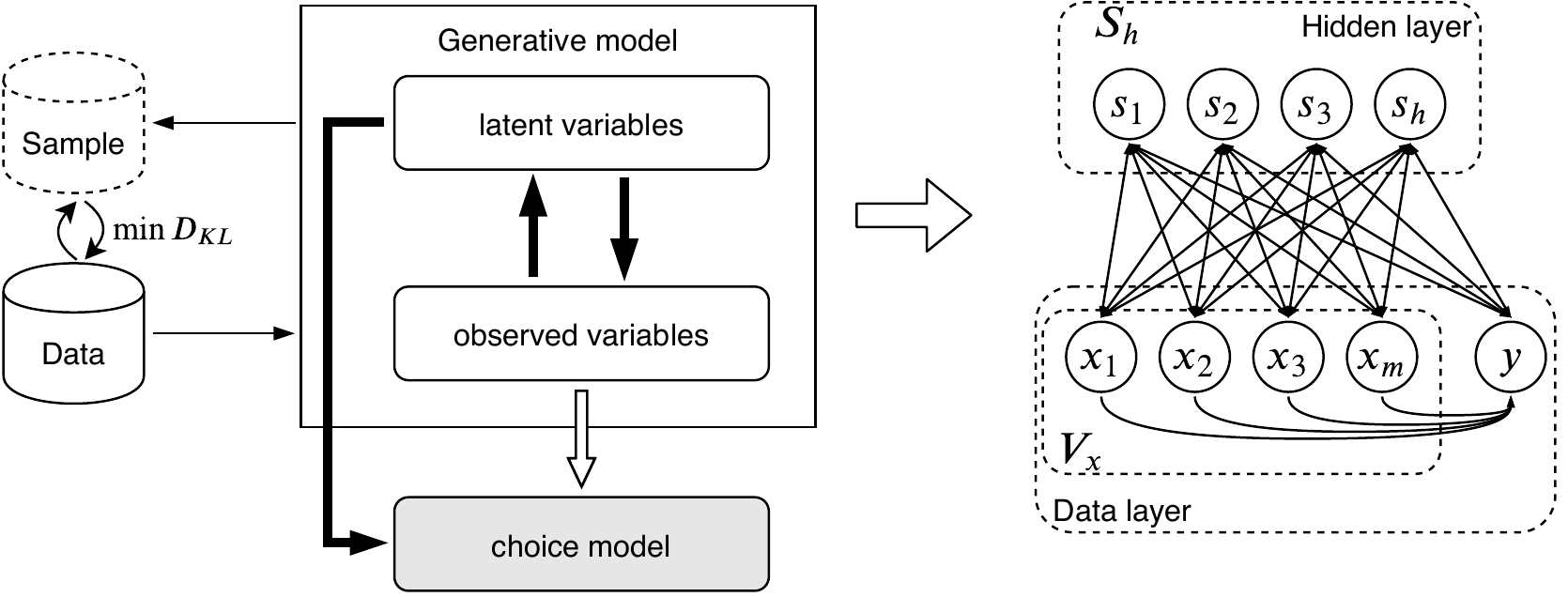}
    \caption{Framework for generative modelling. }
    \label{fig:gen_model}
\end{figure}

Our contributions are as follows: 
(i) A novel framework for capturing and extracting properties of information heterogeneity in travel behaviour models (\cref{fig:gen_model}).
(ii) We show that generative modelling can be framed as an abstraction of rational inattention theory. Specifically, the learning and optimization process of a generative model emulates the internal information processing constraints of decision making.
(iii) Demonstration of a data-driven modelling approach that exploits start-of-the-art deep learning techniques. A generative model architecture is described in the methodology. 
(iv) Discussion on the interpretation of generative learning on discrete choice analysis.
(v) We provide new insights into sensitivity analysis of econometric parameters through a travel behaviour case study.

The remainder of the paper is organized as follows: \cref{sec: Information theory in behaviour models} introduces preliminary concepts related to information theory in choice modelling and discusses existing literature on rational inattention behaviour theory. 
\cref{sec: Methodology} describes the generative model framework and estimation methodology. 
In \cref{sec: case study}, a case study example on a trip-based travel behaviour analysis is shown and we demonstrate how the results explain information heterogeneity in the data.
\cref{sec:conclusion} provides a brief discussion on the results, conclusion and suggestions for future research.

\section{Information theory in behavioural models}
\label{sec: Information theory in behaviour models}
In this section, we introduce several preliminary concepts that relates to our work by beginning with the connection between rational inattention behaviour and information theory in the context of generative modelling.

\subsection{Rational inattention behaviour}
Rational inattention presents a behavioural scenario where individuals' choice influences are based on Shannon's mutual information that measures uncertainties between an exploitative and exploratory choice process. 
Specifically, it frames the choice problem on observations as well as information processing constraints similar to that of a communication channel with finite Shannon capacity \citep{sims2003}.
By representing information processing constraints, it accounts for the natural deviations in econometric behaviour \citep{sims2003,sims2010}.
This concept stems from the same principles of neuroscience where behaviour learning and perceptual inference can be explained through information theory and statistical physics \citep{friston2006}.
Using modern deep learning techniques, one can construct a rational inattentive learning model using an artificial neural network to provide a principled way of analyzing travel behaviour patterns from large scale datasets.

As a simple generalization, information processing constraints across choice preferences can be represented by an unknown distribution of random utility shocks according to Ellsberg's paradox which showed that individuals systematically violate utility theory by being adverse to ambiguity \citep{ellsberg1961}. 
Consider a case where an individual is faced with two options in a choice set when the expected utilities are identical for both options. 
In utility theory, both options will be chosen at equal probabilities, whereas in rational inattention, the individual chooses the option that maximizes entropy (attention).
This decomposition accounts for the prediction error under different protocols as well as it resembles exploratory choice behaviour (i.e. prospect theory) \citep{kahnemantversky1979}.
For instance, when the differences in utility between two travel modes do not differ, travellers would try new options, in relation to increased risk.

Existing studies on rational inattention in choice modelling research stems from the findings that this behaviour can be generalized in an MNL model \citep{matvejkamckay2015}.
However, they have mostly focused on static models, as dynamic rational inattention models are difficult to solve and may be intractable using conventional methods \citep{steineretal2017}.
The value of adding information processing constraints have suggested well-defined similarities with macroeconomic behaviour theory \citep{sims2003}.
Recently, rational inattention has become a particularly appealing approach to modelling choice behaviour. For instance, \cite{matvejkamckay2015} described the implication of information availability on consumer choice selection behaviour using a rational inattention model. 
In a combined location and mode choice model, \cite{teyeetal2017} used a method of entropy maximization in a non-linear mixed integer program subject to available information constraints.
\cite{leard2018} investigated consumer inattention correlation with willingness-to-pay for fuel consumption. 
Recently, rational inattention has been found to work well in time variability problems in travel demand forecasting \citep{fosgeraujiang2019}. 
The theory of rational inattention seeks to endogenize the imperfect awareness about the circumstances \citep{sims2010}. 
The decision maker selects pieces of information that are most relevant for his or her utility and ignores the rest, so long as the information cost can be accounted for in the model. 

\subsection{Information theory}
In this section, we explore some key properties of information theory in the context of behavioural modelling.
Information theory has been used to provide insights into the non-rational behavioural choice, and it was shown to be equivalent to random utility maximization MNL model \citep{anas1983}.
An information theoretic model can also be used as a tool for generating new predictions beyond MNL restrictions, subject to available information \citep{anas1983}. 
Recent studies have also shown that this is also functionally equivalent to an additive random utility maximization problem in rational inattention behaviour models and several well-known decision problems can be reasonably represented, e.g. Prospect Theory and Regret Theory \citep{kahnemantversky1979,matvejkamckay2015}.
The measure of information heterogeneity is closely related to non-normative representation, involving Shannon entropy \citep{fosgerauetal2017}.
Expected utility representation may not be sufficient in providing the proper specification for these decision problems as individuals may perceive choice probabilities with different levels of uncertainty.
Decisions under uncertainty can be interpreted in a simple way by correcting for information processing constraints in the utility specification.

\subsubsection*{Energy}
Assuming a bi-directional system with an observed and an unobserved (latent) states, the level of uncertainty of a state configuration of the system with observed \(X\) and latent \(S\) random variable is a function of energy \(E(x,s)\) of the state proportional to the joint probability \(p(X=x,S=s)\) or \(p(x,s)\):
\begin{equation}
    \label{energy_1}
    p(X=x,S=s) = \frac{1}{Z}e^{-E(x,s)},
\end{equation}

where \(Z=\sum_{x,s}e^{-E(x,s)}\) is the normalization function so that \(\sum_{x,s}p(x,s)=1\). 
Due to the logarithmic function, energy decreases monotonically as the probability increases.
Imposing monotonicity allows the model estimates to be more interpretable and tractable.
An event with high energy will have a lower probability of occurrence (individuals will tend to avoid this state).
An event with low energy will always be within the expectation of the individual, thus having higher probability \citep{ullah1996}.

\subsubsection*{Mutual information}
Mutual information allows us to identify general nonlinear dependencies by measuring the amount of information processed by the individual, i.e. how far two random variables are from being independent.
Given two random variables \(X\) and \(S\), let \((X,S)\sim p(x,s)\), the mutual information \(I(X,S)\) can be written in the form:
\begin{equation}
    \label{mutual_information}
    I(X,S) = \sum_{x,s}p(x,s)\log\frac{p(x,s)}{p(x)p(s)} = \sum_{x,s}p(x,s)\log\frac{p(x|s)}{p(x)} = H(X) - H(X|S)
\end{equation}

It can be interpreted as the decrease in uncertainty of X given S, where H(X) and H(X|S) are the entropy and the conditional entropy respectively:
\begin{equation}
    H(X) = -\sum_s\sum_x p(x,s)\log p(x) = -\sum_s p(s|x) \sum_x p(x) \log p(x) = -\sum_x p(x) \log p(x)
\end{equation}

Mutual information is symmetric $I(X,S) = I(S,X)$ and it is non-negative, $I(X,S)\ge 0$ and it is zero if and only if \(X\) and \(S\) are \textit{independent} (with respect to the model identification process).
Hence, the mutual information shown in \cref{mutual_information} is equivalent to finding the expected energy difference between the data generating distribution and the true distribution obtained from the data.

\subsection*{Kullback-Leibler divergence}
The Kullback-Leibler (KL) divergence or the relative entropy measures the `distance' between two distribution, \(p\) and \(q\) \citep{ullah1996}.
The KL divergence of \(q\) from \(p\) is \(D_{KL}(q||p)\), when \(q=p\) then \(D_{KL}=0\).
The mutual information, using the example above, can be defined as the divergence of the joint distribution from the product of marginals:
\begin{equation}
    I(X,S) = D_{KL}(p(x,s)||p(x)p(s)) \geq 0
\end{equation}

Thus in practice, we can consider the hypothesis \(H_0:D_{KL}=0\) against \(H_1:D_{KL}\neq 0\) as a test for \textit{independence} between two random variables \citep{ullah1996}.
To put it in a different perspective, if we can define a framework where the latent variables interact with the observed variables by a correlation matrix, then the mean and variance of the matrices indicate how much information heterogeneity is present in the data describing the population.

\section{Methodology}
\label{sec: Methodology}
We propose a generative model framework that extends rational inattentive behaviour in discrete choice, interpreting it as an \textit{optimization process} rather than a structural model specification.
We differentiate our work from the generalized entropy function described in \cite{fosgerauetal2017} by framing non-normative behaviour as a learning model -- allowing for random perturbations to be data-driven.
Under this framework, the estimation of a generative model assumes to emulate information processing constraints in rational inattention behaviour and identifies observed and latent variable interactions through a neural network interface. 
The correlation between random decision and information priors are reflected through the estimated latent variable parameters.
We use a Restricted Boltzmann Machine (RBM) learning algorithm as an example to estimate the generative model parameters.
Other forms of generative model algorithms (e.g. Autoencoders, GANs\footnote{GAN: Generative Adversarial Networks}, DBNs\footnote{DBN: Deep Belief Nets} \citep{goodfellowetal2016}) can similarly be used.
Another simpler form of generative modelling is principal component analysis (PCA).
However, PCA has severe limitations as it cannot handle complex non-linear relations in the data \citep{hintonsalakhutdinov2006}.
We focus on the RBM learning algorithm as we would show that it is an approximation to a rational inattention information processing with similarities to an error components model.
The error components control for the heterogeneity in the observed utility and variances in the unobserved utility, where the unobserved utility is represented by an entropy function.

\subsection{Proposed generative model framework}
The generative model framework is a tri-partite RBM with a data layer \(\mathcal{D}\) representing the set of observed variables \(\mathcal{D}=\{x_1,x_2,...x_m, y\}\) including a dependent variable \(y\) and a hidden layer \(\mathcal{S}\) representing the set latent variables \(\mathcal{S}=\{s_1,s_2,...,s_h\}\) (see \cref{fig:gen_model}).
The generative model can be framed as a fully connected tri-partite graph \(\mathcal{G}=(\mathcal{V},\mathcal{E})\) where \(\mathcal{D,S}\in\mathcal{V}\) is the set of graph nodes and \(\mathcal{E}\) are the graph edges.
The nodes from \(\mathcal{V}_x=\{x_1,x_2,...,x_M\}\) are connected to \(\mathcal{V}_y=\{y\}\) by edge subset \(\mathcal{E}_{xy}\), representing the choice model explanatory variable coefficients.
The edges between \(\mathcal{S}\) and \(\mathcal{D}\) are the correlation matrix between the latent and observed variables. 
Decision level heterogeneity is represented by the edge subset \(\mathcal{E}_{hy}\).
The algorithm focuses on generating synthetic data using a blocked Gibbs sampling protocol, alternating between observed and latent variable samples from the joint distribution conditioned on the previous step.
A non-zero valued covariance matrix represents the level of information heterogeneity captured in the data.
A zero covariance matrix indicates that the observed explanatory variables captures all the taste variations and assumes a fully homogeneous population.
The observed data can be inferred by sampling from the generative model probability distribution.
By minimizing the KL divergence between the observed and generated data, we learn the parameters of the correlation matrix between the observed and latent variables.
When the generated data have matched the observations, the underlying priors are assumed to have encoded the information heterogeneity of the population and can be represented in the choice model.

\subsection{Model specification}
The RBM architecture was designed as an efficient feature descriptor that progressively trains a fully connected non-linear model structure \citep{hintonetal2006}.
The interactions between the two parallel components capture the information about the heterogeneity present between hypotheses.
Each latent variable represents a specific state encoded as distributed binary patterns.\footnote{Distributed binary patterns are commonly used in digital signal encoding. 
For example of a pattern: \(\mathbf{s}=\{0,1,0,0\}\) or \(\{1,1,0,1\}\). We make the analogy to digital encoding to refer to choice behaviour perceptions. 
A latent variable model with \(N\) elements can represent up to \(2^N-1\) different behaviour perceptions. The Boltzmann architecture uses this representation with a stochastic sampling algorithm to learn the model parameters.
Other forms such as multinomial discrete vectors or multivariate normal can also be used as possible encoding patterns, but binary encodings are the most straightforward method to simplify model inference.}
The different combinations of latent variables form the complex behavioural activity patterns and are inferred through sampling from the posterior.
Similar to a random utility specification, we start with a scalar energy value describing the joint configuration of observed explanatory variables, dependent choice variable and latent variables:
\begin{equation}
    E(\mathbf{x},\mathbf{s},y) = -\mathbf{x}\boldsymbol{\beta}y - \mathbf{xWs} - \mathbf{sW'}y - \mathbf{dx} - \mathbf{c}y - \boldsymbol{\alpha}\mathbf{s}
\end{equation}

The energy function is parameterized by a set of coefficients \(\phi=\{\boldsymbol{\beta}, \mathbf{d}, \mathbf{c}, \boldsymbol{\alpha},\mathbf{W}\), \(\mathbf{W}'\}\), where \(\boldsymbol{\beta}\) are the choice model coefficients and \(\mathbf{d}, \mathbf{c}, \boldsymbol{\alpha}\) are the constants of the observed explanatory, dependent and latent variables respectively.
\(\mathbf{W}\) and \(\mathbf{W}'\) are the parameters matrices representing the information heterogeneity captured by the latent variables given the observed explanatory and dependent variables. 
\(\mathbf{y}\) is a discrete dependent variable representing the choice alternatives, e.g. \(y=\{1,0,0\}^\top,\{0,1,0\}^\top\) or \(\{0,0,1\}^\top\) representing a selected alternative.
\(\mathbf{x}\) is a vector of observed explanatory variables either as discrete or continuous values.
Multiple discrete and continuous dependent values can also be used as the output \citep{wongfarooq2019}.
\(\mathbf{s}\) is a vector of stochastic binary variables.
Given that the non-latent variable terms can be factorized out, the posterior over the latent variables is as follows:
\begin{equation}
    \label{eq:factorised_posterior}
    p(\mathbf{s}|\mathbf{x},y) \propto \prod_h p(s_h|\mathbf{x},y) = \prod_h \exp({\mathbf{xW}_h s_h + s_h\mathbf{W}'_h y + \alpha_h s_h})\\
\end{equation}

Using the aforementioned energy function \(E(\mathbf{x},\mathbf{s},y)\) allows the conditional to be factorized.
Defining the normalizing constant as the sum of the binary configurations, we obtain the \textit{normalized} probability density function for each latent variable \(s_h\):
\begin{align}
    p(s_h=1|\mathbf{x},y) &= \frac{\tilde{p}(s_h=1|\mathbf{x})}{\tilde{p}(s_h=0|\mathbf{x}) + \tilde{p}(s_h=1|\mathbf{x})}\\
                        &=\frac{1}{1+\exp(-((\mathbf{xW})_h + \mathbf{W}'_h y + \alpha_h))}
\end{align}

The objective is to optimize the model parameters such that a sample \(\tilde{\mathcal{D}} = \{\tilde{x}_2,\tilde{x}_2,...,\tilde{x}_m,\tilde{y}\}\) is generated with a distribution as close to the data distribution \(\mathcal{D}\).
Computing the energy over the data layer \(E(\mathcal{D})\) corresponds to the expected energy of the model minus the entropy:
\begin{equation}
    E(\mathcal{D}) = \sum_s p(s|\mathcal{D}) E(\mathbf{x},\mathbf{s},y) - H(S)
\end{equation}
which can be simplified into the form:
\begin{align}
    E(\mathcal{D}) &= -\log \sum_s e^{- E(\mathbf{x},\mathbf{s},y)}\\
                   &= -\mathbf{x}\boldsymbol{\beta}y - \mathbf{dx} - \mathbf{c}y -\log\Bigg(\sum_s \Big(\exp(\mathbf{xWs} + \mathbf{sV}y + \boldsymbol{\alpha}\mathbf{s})\Big)\Bigg)\\
                   &= -\mathbf{x}\boldsymbol{\beta}y - \mathbf{dx} - \mathbf{c}y -\log\Bigg(\prod_h\Big(\sum_{s_h\in\{0,1\}} (\exp((\mathbf{xW})_h s_h + s_h\mathbf{W}'_hy + \alpha_hs_h))\Big)\Bigg)\\
                   &= -(\boldsymbol{\beta}y + \mathbf{d})\mathbf{x}  - \mathbf{c}y -\sum_{h}\log\Bigg(1+\exp((\mathbf{xW})_h + \mathbf{W}'_hy + \boldsymbol{\alpha})\Bigg)
                   \label{eq:energy_minus_entropy}
\end{align}

\cref{eq:energy_minus_entropy} is a direct interpretation of the generalized entropy formulation for discrete choice \citep{anas1983,fosgerauetal2017}.
The coefficients \((\boldsymbol{\beta}y + \mathbf{d})\) stand for the unknown parameters of the explanatory variables for each alternative \(y\) and for the generative model respectively.
Increasing \(\boldsymbol{\beta}\) decreases the energy over the data generating distribution conditioned on a choice alternative, while increasing \(\mathbf{d}\) decreasing the energy over all data generating configurations.
\(\mathbf{c}\) represents the alternative specific constants and \(\sum_{h}\log(1+\exp(\mathbf{xW}_h + \mathbf{W}'_hy + \boldsymbol{\alpha}))\) is the flexible error component generator given a specific input configuration of observed \(\mathbf{x}\) and \(y\) with a constant \(\boldsymbol{\alpha}\).
If this term is near zero, The expected energy function is equivalent to a utility function in a random utility maximizing (RUM) model.
By definition, the probability of \(\tilde{\mathcal{D}}\) being generated is the Boltzmann distribution with energy \(E(\mathcal{D})\):
\begin{equation}
    p(\tilde{\mathcal{D}}) = \frac{1}{Z'}e^{-E(\mathcal{D})}
\end{equation}

The computation of the marginal \(Z'=\sum_{\mathcal{D}'} e^{-E(\mathcal{D}')}\), which sums over an exponential number of possible configurations of the data vector, becomes difficult as we increase the number of explanatory variables.

\subsection{Objective function formulation}
Our proposed framework addresses the estimation problem for a highly non-linear and non-closed form function using variational inference.
We select from a family of distributions that produce an \textit{approximate} posterior distribution.
The specification of the posterior distributions is obtained from data accumulation during the learning phase.
If we restrict the family of distributions that are tractable and can be factorized over each variable in \(Z\), the problem of simulation-based estimation becomes significantly simpler.
For the sake of clarity, we omit the parameter terms \(\phi\) in the equations below.
First, we consider \(p(\tilde{\mathcal{D}})\) in terms of energy and the joint probability as follows:
\begin{equation}
    \label{log_p_energy}
    p(\tilde{\mathcal{D}}) = \sum_s p(\mathcal{D},\mathbf{s}) = \frac{e^{-E(\mathcal{D}})}{\sum_{\mathcal{D}'} e^{-E(\mathcal{D'}})}\\
\end{equation}

We can map the energy of the observed part as a function of the total system energy in a formulation similar to \cref{energy_1} by defining \(E(\mathcal{D}) = -\log\sum_s e^{-E(\mathcal{D},\mathbf{s})}\).
The posterior over the latent variables as a function of energy using Bayes rule, \(p(a|b)=p(a,b)/p(b)\) results in a Boltzmann probability function over the joint distribution, which reveals the similarities to an MNL model:
\begin{equation}
    \label{posterior_as_bd}
    p(s|\mathcal{D}) = \frac{p(\mathcal{D},s)}{\sum_{s'}p(\mathcal{D},s')} = \frac{e^{-E(\mathcal{D},\mathbf{s})}}{\sum_{\mathbf{s}'}e^{-E(\mathcal{D},\mathbf{s}')}}
\end{equation}

If we take the expected values with respect to the posterior on (Eq. \ref{log_p_energy}), the uncertainty of choice can be expressed in terms of expected energy and entropy denoted as the evidence lower bound \(\mathcal{L}\):
\begin{equation}
    \label{F_U_H}
    \mathcal{L} = \underbrace{- \underbrace{\left[\sum_s p(s|\mathcal{D})\right]}_{=1} \log p(\tilde{\mathcal{D}})}_{\textrm{uncertainty}} = \underbrace{\sum_s p(s|\mathcal{D}) E(\mathcal{D})}_{\textrm{expected energy}} - \underbrace{\Bigg(-\sum_s p(s|\mathcal{D}) \log p(s|\mathcal{D})\Bigg)}_{\textrm{entropy gain}}
\end{equation}

In \cref{F_U_H}, a rational inattentive based choice can be framed as the information difference between the expected energy and the entropy gain. 
The first term on the right of \cref{F_U_H} denotes the individuals' behaviour towards prior expectations about the choice. 
The second term is the entropy and it can be viewed as the information processing constraints in a rational inattentive model or a penalty for low energies.
It ensures that the generative model produces low uncertainty values for inputs with high probability in the true data distribution and high uncertainties for all other inputs \citep{ranzatoetal2007}.
minimizing uncertainty implies both utility maximizing and entropy seeking behaviour. 
Computing the evidence \(\log p(\tilde{\mathcal{D}})\) is intractable, but we can use the posterior \(p(s|\mathcal{D})\) to evaluate the marginal log likelihood \citep{kingmawelling2013}. 

In many cases, computing the posterior \(p(s|\mathcal{D})\) may be difficult when the distribution is complex, as we require an integral over all configurations of latent variables to find the marginal or denominator in \cref{posterior_as_bd}. 
The primary motivation of defining the problem as \textit{variational inference} is that we can approximate the posterior distribution using a tractable arbitrary distribution \(q(s)\) \citep{bleietal2017}.
In the estimation procedure, we find the parameters that make \(q\) as close as possible to the posterior by minimizing \(\mathcal{L}\) where \(q\) is the approximating distribution, then we have:
\begin{equation}
    \label{pre_kl}
    -\log p(\tilde{\mathcal{D}}) = \mathbb{E}_{q(s)}\left[ E(\mathcal{D}) -(-\log p(s|\mathcal{D}))\right] 
\end{equation}

To show that the proposed distribution \(q(s)\) can be used to approximate \(p(s|\mathcal{D})\), we compute the marginal loglikelihood over \(q(s)\) to minimize the KL divergence of \(q(s)\) from \(p(s|\mathcal{D})\):
\begin{align}
    -\left[\sum_s q(s)\right]\log p(\tilde{\mathcal{D}}) &= \mathbb{E}_{q(s)}\left[E(\mathcal{D})\right] -(- \mathbb{E}_{q(s)}\left[\log p(s|\mathcal{D})\right]) \\
               &= \sum_s q(s) E(\mathcal{D}) + \sum_s q(s) \Bigg( \log p(s|\mathcal{D}) + \log \frac{q(s)}{q(s)}\Bigg)\\
               &= \sum_s q(s) E(\mathcal{D}) + \sum_s q(s)\log q(s) - \sum_s q(s)\log\frac{q(s)}{p(s|\mathcal{D})}\\
               &= \underbrace{\sum_s q(s) E(\mathcal{D}) - H_q(S)}_{\textrm{variational free energy } F_q(\mathcal{D})} - D_{KL}(q(s)||p(s|\mathcal{D}))
\end{align}

Using the fact that the KL divergence cannot be negative, we get the lower bound on the model evidence and we define the \textit{variational} free energy \(F_q(\mathcal{D})\) as:
\begin{equation}
    \label{min_var_energy}
    F_q(\mathcal{D}) = \mathcal{L} + D_{KL}(q(s)||p(s|\mathcal{D})) \geq \mathcal{L}    
\end{equation}

The intuition from \cref{min_var_energy} is that minimizing the variational energy has the same outcome as minimizing \(D_{KL}(q(s)||p(s|\mathcal{D}))\). 
The bound is exact if \(D_{KL}(q(s)||p(s|\mathcal{D}))\) term is zero, which would happen if \(q(s)\) matches \(p(s|x)\) perfectly.
Therefore, following the gradient of \(F_q(\mathcal{D})\) yields the optimal solution for \(q(s)\).
Another equivalent form of variational free energy can be derived by transforming the marginal into the conditional likelihood:
\begin{equation}
    \label{conditional_elbo}
    F_q(\mathcal{D}) = -\log p(\mathcal{D}|s) + D_{KL}(q(s)||p(s)) 
\end{equation}

In \cref{conditional_elbo}, the objective function can be optimized through assigning specific priors over the generative model then measuring how well the priors represent the observations.
More generally, minimizing \(F_q(\mathcal{D})\) together with the KL divergence is a good substitute for minimizing the log-likelihood function \citep{ranzatoetal2007}.
The first and second terms on the right-hand side are known as the fit and complexity respectively in Bayesian statistics.
The first term defines the accuracy of the data generating model.
If we presume that \(p(s)\) is a complex model (real-world representation, intricate correlation between behaviour and choices, etc.), then the complexity tells us how much capacity is required for the (non-trivial) approximator \(q(s)\) to match the empirical distribution.
The variational energy can be used to determine the strength of non-linear interactions between components in a model.
Minimization of variational energy provides consistent and reproducible models, equivalent to maximum likelihood estimation.
We can establish the choice model by interpreting the data generating probabilities of a given data vector as the individuals' information heterogeneity by minimizing the variational lower bound. 
The objective cost function now becomes selecting the model parameters such that:
\begin{equation}
    \theta^* = \operatorname*{arg\,min}_\theta \{D_{KL}(q(s)||p(s|\mathcal{D}))\}
\end{equation}

In the proposed generative model, we are interested in evaluating large numbers of non-linear latent variables which belongs to a family of extreme valued distributions parameterized by latent variable parameters \(\theta=\{\mathbf{d},\boldsymbol{\alpha},\mathbf{W}\), \(\mathbf{W}'\}\).
The primary assumption is that the approximating distribution \(q(s)\) can be factorized, such that it gives a tractable form:

\begin{equation}
    \label{conditional_prob}
    q(s) = \prod_h q(s_h;\theta) \approx \prod_h p(s_h|\mathcal{D})
\end{equation}

This form allows the generative model to produce distributions with sharper boundaries over conventional mixture models.
Using this specification, model variance can be increased or decreased with the number of activated latent variables.

\subsection{Parameter estimation}
We formalize the model learning as minimizing KL divergence given some observed data \(\{\mathcal{D}^n\}_1^\infty\). 
The key advantage of this is that we can incorporate the differences between individual's actual behaviour and mean population behaviour effectively in the objective function.
The parameter update rule for a generative model is obtained by implementing a stochastic gradient descent on the variational free energy function, updating the weights of the coefficients between latent and observed variables according to the sampling states.
Consequently, the gradients with respect to the parameters are as follows:
\begin{equation}
    \label{cdn}
    \mathbb{E}_q\left[\frac{\partial}{\partial\theta}\log p(\tilde{\mathcal{D}})\right] = \sum_{n=1}^\infty\frac{\partial D_{KL}(q(s)||p(s|\mathcal{D}^n))}{\partial\theta} \approx\frac{\partial}{\partial\theta}E(\mathcal{D}^{(1)},s) - \mathbb{E}[\frac{\partial}{\partial\theta}E(\mathcal{D}^n,s)],
\end{equation}
where the expectation is over \(\tilde{\mathcal{D}}\sim p(\tilde{\mathcal{D}})\).
The learning algorithm is based on a Gibbs chain starting at an initial sample \(\mathcal{D}^{(1)}\) from the data distribution and converging to the RBM data generating distribution after performing alternating blocked Gibbs sampling between the latent and observed variables.
A naive implementation of this learning algorithm would require simulating the Gibbs sampler to equilibrium after every model update before drawing a new set of observations from the data.
Sampling from the generative model to produce \(\mathcal{D}^{(1)},...,\mathcal{D}^n\) with \(n\leq 10\) and updating the model parameters between each iteration has been suggested as a optimal tradeoff between fast estimation without loss in generality or stability \citep{hintonetal2006}.
The first term on the right-hand side of \cref{cdn} is the derivative of the energy function w.r.t the initial Gibbs samples and the second term corresponds to the gradient of the energy function after \(n\) steps.

Our proposed modification to the RBM learning algorithm uses a hybrid generative learning and maximum utility estimation. 
Rather than focusing solely on the optimization of the generative component, we also try to maximize the accuracy of our choice model given the data and generative samples.
After each generative learning step, we update the choice model coefficients by performing maximum likelihood on the \textit{conditional} using the choice alternative as the dependent variable.
Next, we sample latent variables from the generative model using the observed explanatory variables as inputs. 
These latent variables are assumed to represent the information heterogeneity that is not captured by the explanatory variables.
Our modification provides integration with discrete choice modelling methods and allows for other hybrid choice model use cases that can be explored in the future.
We specify the conditional logit model using observed and latent variables as follows:
\begin{equation}
    p(y_j=1|\mathbf{x,s}';\beta_j,c_j) = \frac{\exp\Big((\beta_j+\mathbf{d})\mathbf{x} + c_j + \sum_h\log(1+\exp((\mathbf{xW})_h + \mathbf{W}'_{hj}+\alpha_h))\Big)}
    {\sum_{j'}\exp\Big((\beta_j+\mathbf{d})\mathbf{x} + c_{j'} + \sum_h\log(1+\exp((\mathbf{xW})_h + \mathbf{W}'_{hj'}+\alpha_h))\Big)},
    \forall \{\mathbf{x},y\} \subseteq \mathcal{D},
    \label{conditional_ouput}
\end{equation}
where there are \(j\) alternatives in the choice variable \(y\).
In this step, only the \(\beta\) coefficient and \(c_j\) alternative specific constants are updated (by maximum likelihood) while keeping the parameters from the generative model unchanged.
Given that parameters \(\mathbf{W}'_{hj}\) and \(\alpha_h\) are estimated from the generative model learning algorithm providing model error correction, the coefficients of the choice model is expected to converge to a non-biased, homogeneous value.
This means that as we improve the precision of the data generation protocol, the choice model can be estimated without systematic errors.

\subsection{Economic interpretation}
The basis for economic interpretation of a generative model is through a combination of individual utility and entropy.
Suppose that an individual will be in one of \(S\) latent decision states, each state has associated with it a configuration of latent variables: \(\{s_1,...,s_h\}\).
These latent variables are related to choice selection strategies, complexity and influence of repeated nature of travel activity choices.
Thus they are interpreted as potential decision strategies.
If in a particular state \(S\) contains all zero elements, then the choice strategy is a purely utility driven one (since latent variable attributes are ignored).
If by contrast, the latent variables are \textit{non-zero}, then one might argue that the individuals used their internal information processing constraints to develop a choice strategy.
These interpretations are similar to the rational inattention model, which were identified as decision strategies characterized by continuously optimizing agents \cite{sims2003}.

We assume some distribution function to describe \(G_j\), an error generating density function that depends on \(\{s_1,...,s_h\}\) for all \(j\) alternatives.
The density \(G_j\) is the distribution of the unobserved heterogeneity on the individuals with similar utilities for each alternative.
It represents the idealistic subjective perception of a particular individual on a specific choice context.
We assume that \(\{s_1,...,s_h\}\in[0,1]\) are extreme value distributed across individuals and decisions:
\begin{equation}
    G_j(s_1,...,s_h) = \prod_h (1+\exp((\mathbf{xW})_h + \mathbf{W}'_{hj}+\alpha_h))^{-1}
\end{equation}

This specification allows a form of energy based models to be generated using entropy as a measure without relying completely on hypothesis-driven utility specifications \citep{train2009}.
As such, from \cref{conditional_ouput}, the generative model specification under a generalized extreme valued function can be derived as follows:
\begin{equation}
    P(y_j) = \frac{Y_jG_j}{\mu G},
\end{equation}

where \(Y_j=e^{\nu_j}\), \(\nu_j=(\beta_j+\mathbf{d})\mathbf{x} + c_j\) and \(G(s_1,...,s_h)=\sum_{j'}Y_{j'}G_{j'}\).
\(G(s_1,...,s_h)\) is non-negative, homogeneous of degree \(\mu\) and function \((s_1,...,s_h)\) is \(\geq 0\), \(G=\infty\) when \(s_\kappa \rightarrow \infty\) for \(\kappa=1,...,h\) and \(\partial^r G / \partial (s_1,...,s_h) \geq 0\) if \(r\) is odd and \(\leq 0\) if \(r\) is even.
Thus, the level of uncertainty in a choice due to information heterogeneity is described using a function calculated on a set of prior weights and latent variables.
The resulting approximate entropy is given as the negative log of the error generating function:
\begin{equation}
    H_j = -\log G_j(s_1,...,s_h) = \sum_h \log (1+\exp((\mathbf{xW})_h + \mathbf{W}'_{hj}+\alpha_h))
\end{equation}

We can expand the model from an MNL specification by substituting \(V_j = \nu_j + H_j\):

\begin{equation}
    P(y_j) = \frac{e^{V_j}}{\sum_{j'}e^{V_{j'}}} = \frac{e^{\nu_j + H_j}}{\sum_{j'}e^{\nu_{j'} + H_{j'}}} 
\end{equation}

where the arguments in \(V_j\) are linearly separated into the observed utility \(\nu_j\) and entropy \(H_j\).
Thus the probability of choosing an alternative is a function of the observed utility, corrected by the information processing cost of the set of alternatives \textit{and} its explanatory variables observed by the decision maker.
An interesting consequence is that \(H_j\) changes at every instance in the variable space i.e. individuals with similar utilities may have different choice distributions.
Furthermore we can conclude that the changes in the decision making policy are influenced in two ways: first, through the direct correlation with the \textit{observed} attributes and second, indirectly through the information processing capacity of the decision maker.
As a result, even though it is impossible to directly measure the result of economic policy changes on the latent variables, we can obtain the mean and variance of the latent parameter distribution to evaluate the information sensitivity with respect to each explanatory variable.

\subsection{Statistics for model evaluation and validation}
One of the ways to obtain statistics for model evaluation and validation is through simulation and hyperparameter search.
Model evaluation can be performed on out-of-sample simulations using adjusted \(R^2\) serves as an equivalent to KL divergence to determine distribution accuracy. 
For evaluation, we fixed some of the input data and use the generative model to produce new data and compare their distribution accuracy.

There are no exact solutions to the number of latent variables required to create an optimal model.
The most commonly used approach is to validate the model by iterative test on various number of latent variables.
We note that validation is only a crude test of performance and there are generally no accepted methods to adequately determine the optimal number of variables.
Several studies in literature have provided so-called `rule of thumb' regarding the number of inputs and layer sizes \citep{alwosheeletal2018}.
However, the optimal number of latent variables used can differ largely between datasets.
Too few latent variables and the model cannot capture the complex structure in the data, too many latent variables may cause overfitting and increases estimation time.

Evaluating the sensitivity of parameters associated with the explanatory variables can be more challenging.
In our experiment, we found that monitoring changes to \(\beta\)-parameters as we increase the number of latent variables work well for sensitivity analysis.
Theoretically, for variables not influenced by information processing constraints, \(\beta\)-parameters should remain consistent.
Otherwise, for variables that are sensitive to information processing constraints, \(\beta\)-parameters would vanish or shrink to a small value as we increase the number of latent variables so that the choice response could not have been derived from that source \citep{sims2003}.
From a macroeconomic perspective, the decision making actions should respond smoothly to external factors and any disturbances or randomness should be distinctive and manifest only from individual's internal information processing constraints \citep{sims2003}.

\subsection{Comparison with supervised neural networks}
The probability distribution in \cref{conditional_ouput} might seem to be equivalent to a single layer neural network (e.g. DNN) with a \textit{softmax} output, we argue that this is not the case.
In a DNN, model parameters are optimized to maximize a \textit{predictive} output \(p(y|\mathbf{x,s})\), which may result in significant overfitting if model is mis-specified or too many hidden units are used.
Using multiple hidden layers may also potentially degrade the model and result in worse performance \citep{heetal2016}.
However, in our approach of using generative modelling, parameters are optimized to reduce information loss by minimizing \(D_{KL}(q||p)\) in the mapping process between observed and latent states, allowing as much of the original data to be reconstructed.
A generative model provides some form of model generalization such that the parameters stay within the range of values that are realistically \textit{representative} of the underlying behaviour, reducing the probability that the model overfits the choice variable.

Since latent variables are stochastic, \(\tilde{\mathcal{D}}\) may not always be generated by the same underlying configuration. 
Likewise, each sample of observed data vector may produce many different configurations of latent variables.
The advantage of using unsupervised learning over supervised likelihood learning methods in discrete choice model is that it provides a flexible, high-level distributed representation and minimizes optimization inefficiencies caused by random initialization \citep{tehetal2003}.
Model optimization uses a greedy learning algorithm to determine the underlying structure that captures the unobserved heterogenities without dependency on aggregate choice samples.
Similar to rational inattention models, entropy in the variational free energy function is the cost of information from sampling from the generative model.

\section{Case study}
\label{sec: case study}

\subsection{Data preparation}
\begin{figure}[!h]
    \centering
    \includegraphics[width=0.8\textwidth]{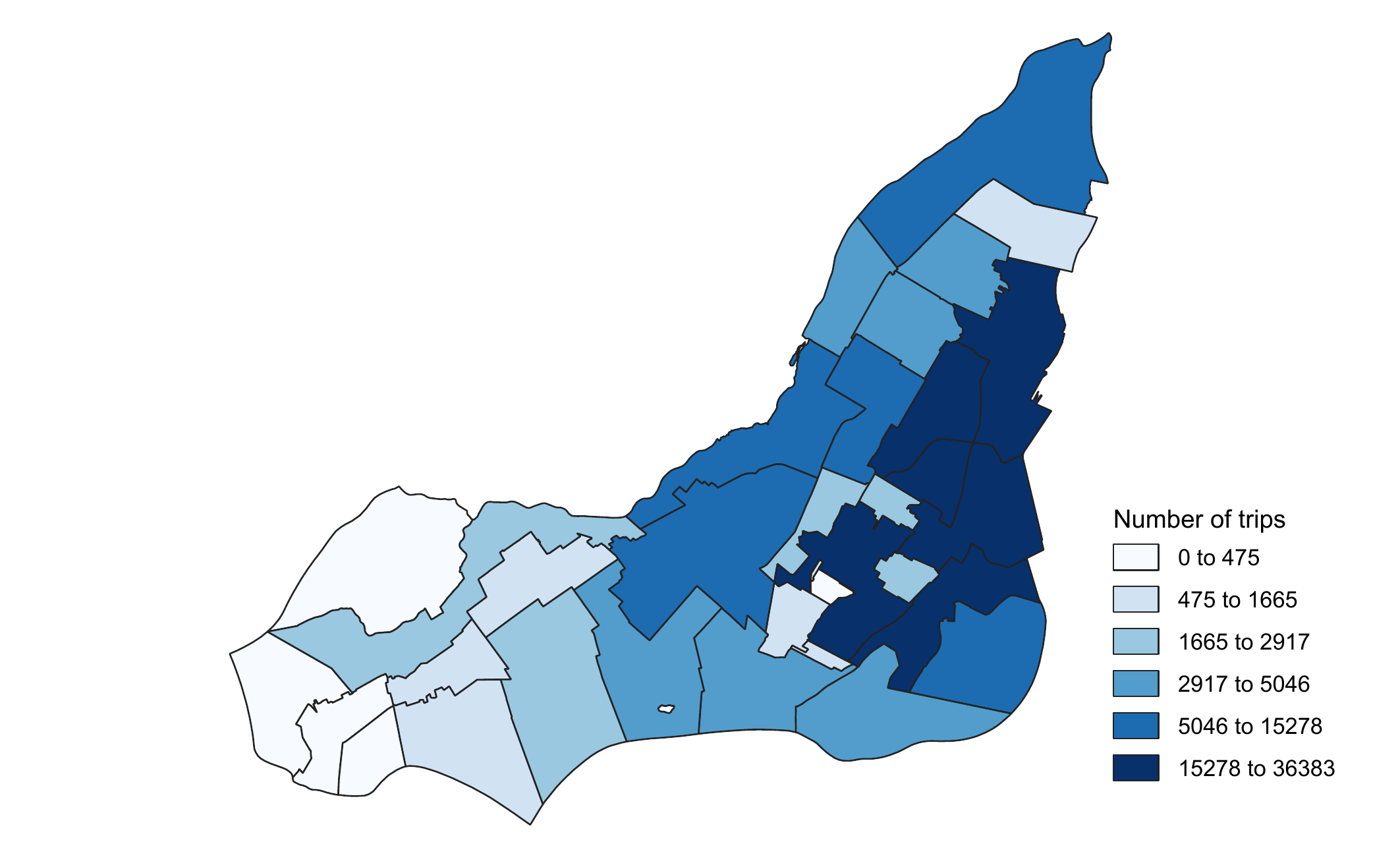}
    \caption{Visualization of number of trip trajectory origin points by city district from the dataset.}
    \label{fig:map1}
\end{figure}

We consider a dataset collected from trip trajectories recorded by respondents from the Greater Montreal Metropolitan Area (\cref{fig:map1}). The data is available as an open dataset provided by the City of Montreal \citep{datamobile2016}.
A total of 293,330 trips observations are available in the dataset and 58,034 trips within these observations have complete travel mode information, purpose and trip characteristics.
We divide the data into two partitions: The first dataset \((\mathcal{D}_{lab},\hspace{0.5em}N_{D}=58,034)\) contains complete (labelled) trip data and is used for model training and validation. 
The second dataset, \((\mathcal{D}_{unlab},\hspace{0.5em}N_{D}=235,296)\) contains incomplete data (unlabelled) and is used for model training, validation, model simulation and analysis. 

For model evaluation, we train a generative model using \(\mathcal{D}_{unlab}\) and \(\mathcal{D}_{lab}\) then we compute the mode choice log likelihood on \(\mathcal{D}_{lab}\) for validation.
The samples are randomly shuffled and split 70:30 for training and validation.
We assume a multinomial extreme valued distribution for categorical observed variables, and log-normal distribution for continuous variables. 
Log-normal is used as the approximation distribution since the continuous data types (speed, distance and duration) follows a positive, right tailed distribution characteristic.
Respective trips of individuals were recorded by self-imputation of their activity for each instance.
Routes of individuals are sampled by GPS traces from their smartphones at frequent intervals.
Speed, distance, activity type, trip duration and trip start location were used as explanatory variables in the estimation. 
The alternatives are: 1:cycling, 2:driving, 3:driving + transit, 4:transit and 5:walk.
Continuous valued variables were normalized to unit standard deviation before model estimation.
A \textit{one-of-j} dummy variable encoding was applied on categorical variables.
A sine/cosine 2D transformation was applied on cyclical continuous values, e.g. time information.

\subsection{Choice model validation}
We present the results of our model validation by assessing the model training performance and analyze the properties of the estimated parameters.
We report the results of our training and validation on model instances with different latent variable sizes: \(S=0\) (standard MNL), \(S=5, S=20, S=35\) and \(S=50\).
In our experiments, we did not notice any significant improvement over 50 latent variables in our model.
To minimize the probability of overfitting in the generative model training, we validate the generative model by monitoring the likelihood loss on the labelled data and select the model parameters at minimum likelihood validation loss.

We used a standard batch stochastic gradient descent (SGD) learning algorithm divided into \(k\) data batches and iterate over \(n\) blocked-Gibbs sampling steps. 
We fixed the hyperparameters for all our experiments to be \(k=16\), \(n=10\), and a learning rate of \(\lambda=0.01\) is used and model parameters are updated in parallel every batch cycle\footnote{The problem of identifying optimal hyperparameters is still not fully understood and it does not provide any useful information with respect to econometric interpretation.
In light of this, we selected these hyperparameters as our baseline for the ease of reproducibility in future work.}.

\begin{figure}[!t]
    \centering
    \includegraphics[width=.8\textwidth]{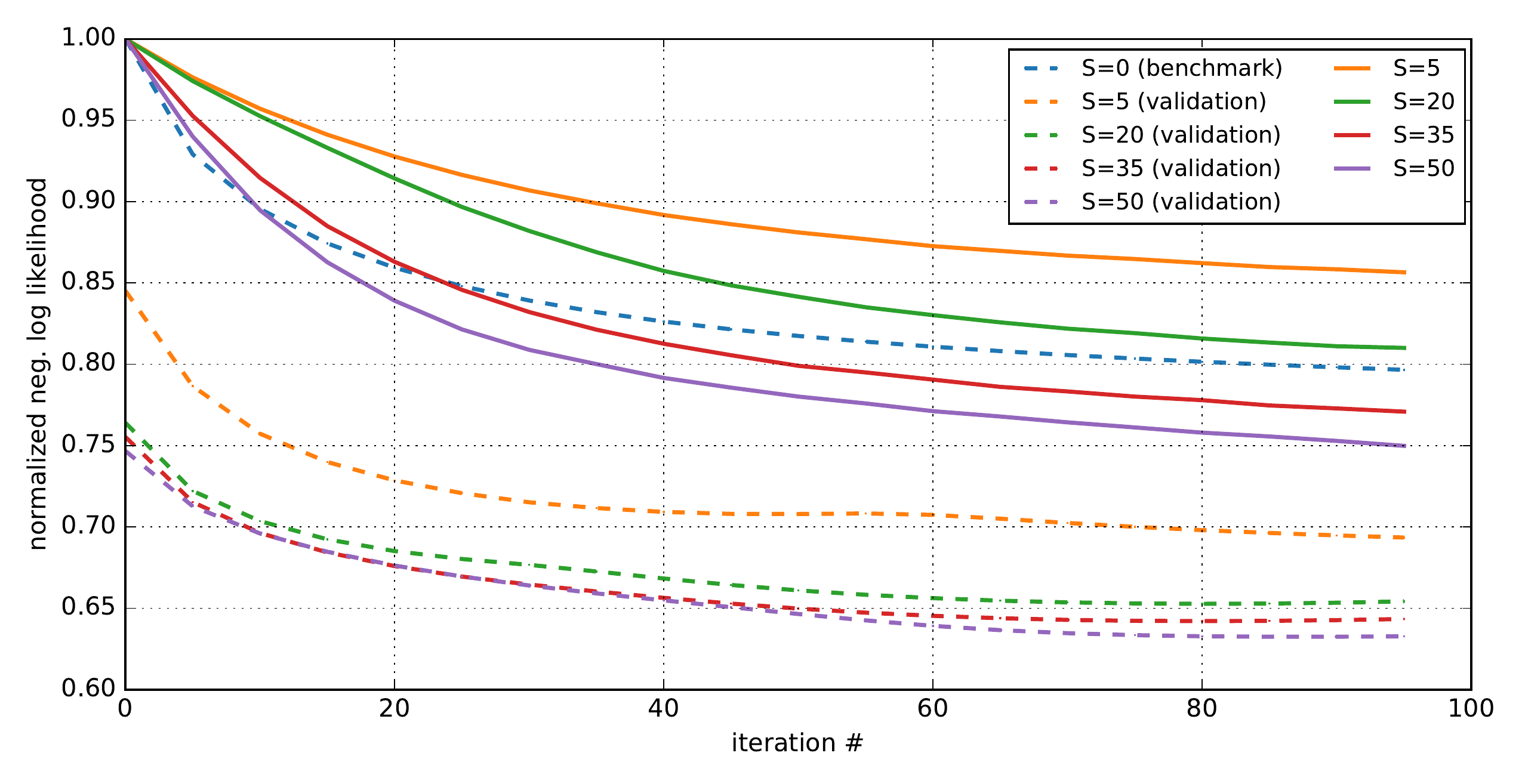}
    \caption{Learning curve of the sample negative loglikelihood from the choice model.}
    \label{fig:curves}
\end{figure}

We monitor validation error by computing the total negative log likelihood of the validation data over the choice model at each iteration.
As observed in the learning curves (\cref{fig:curves}), the model estimation process is stable and converges gradually without overfitting.
At \(S=50\), the model achieved the best overall performance in terms of validation log likelihood.
However, the relative gain in performance decreases as we increase the number of latent variables.
We hypothesize that there is a maximum bound to the effective possible number of latent variables to represent unobserved variations in the data.
This limit can be raised if a greater variation in data is used, i.e., data from different sources or over longer collection time frame.
Note that this analysis is not a test for the `best' mode -- our primary objective is to understand the sensitivity of econometric parameters when a generative learning model is used to account for information heterogeneity.
The loglikelihood decreases rapidly for the first 20 iterations, then plateaued as it reached 100 iterations.
Estimation time for each model instance was less than an 1 hour running our code on a GPU hardware.

\begin{figure}[!t]
    \centering
    \includegraphics[width=.75\textwidth]{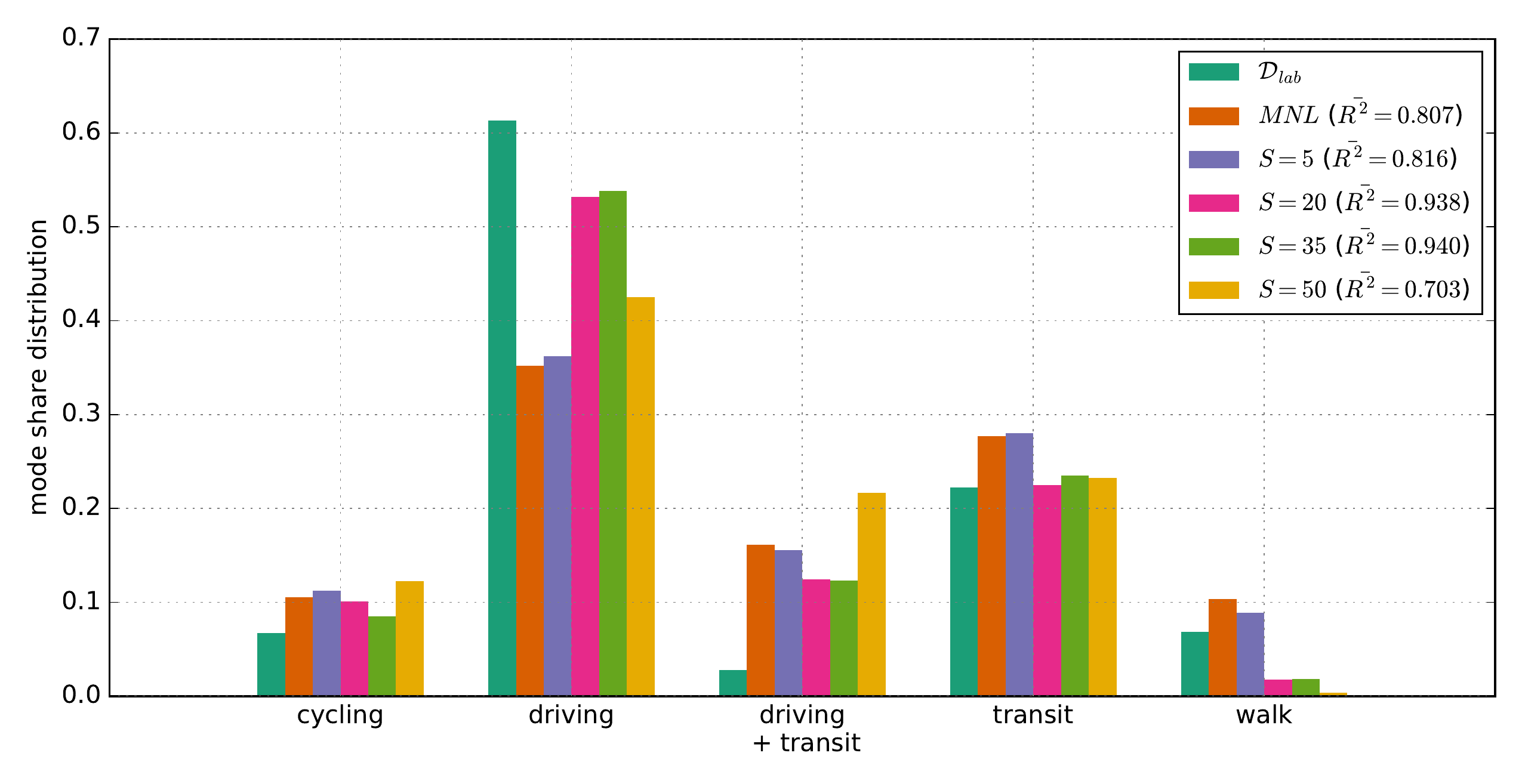}
    \caption{Mode share forecast.}
    \label{fig:sample_mode}
\end{figure}

Model performance is evaluated by comparing the adjusted squared correlation \(\bar{R^2}\) statistical fit.
\cref{fig:sample_mode} shows the mode share distribution of the model validation.
For the baseline model we obtained a \(\bar{R^2}\) value of 0.807
We obtained a \(\bar{R^2}\) value of 0.940, representing a 15\% increase in relative predictive performance.
The nominal trend shows that distribution accuracy increases with an increase in number of latent variables.
At \(S=50\), performance drops slightly compared to \(S=35\) indicating that the performance does not increase asymptotically with the number of latent variables. 
Nevertheless, the results show that the model can be estimated with high accuracy, using KL divergence over maximum likelihood as the objective function.
In this example, the models do not consistently predict the  \textit{driving+transit} and \textit{walking} alternatives probabilities.
One explanation can be attributed to the low observation counts of these two alternatives.
Another possible explanation is that \textit{driving+transit} and \textit{walking} trips have a low correlation with the observed explanatory variables. 

\subsection{Latent variable analysis}
\label{sec: latent variable analysis}
To understand the representational value of latent variables, we analyze their sparse-overcomplete properties \citep{ranzatoetal2007}.
Sparse-overcomplete representation a situation when a large number of latent variables are estimated while only a small number of them are non-zero \citep{ranzatoetal2007}.
It is a practical constraint that allows for more efficient use of latent variables and more flexibility in handling complex correlations which results in a better approximation of the statistical distribution of the data.
Sparse representation has two main advantages in generative modelling \citep{ranzatoetal2008,glorotetal2011}.
The first advantage is that the model will be able to control the dimensionality of representation given a set of inputs, avoiding the overfitting problem.
The second advantage in the context of travel behaviour model inference is that the resulting representation is more likely to be linearly separable, decreasing the complexity in the model even though more parameters are estimated.
This means that even with a large number of latent variables, sparse distribution of parameters would constraint the model to learn distributions which are most statistically significant in reproducing the original data.

\begin{figure}[!t]
    \centering
    \includegraphics[width=\textwidth]{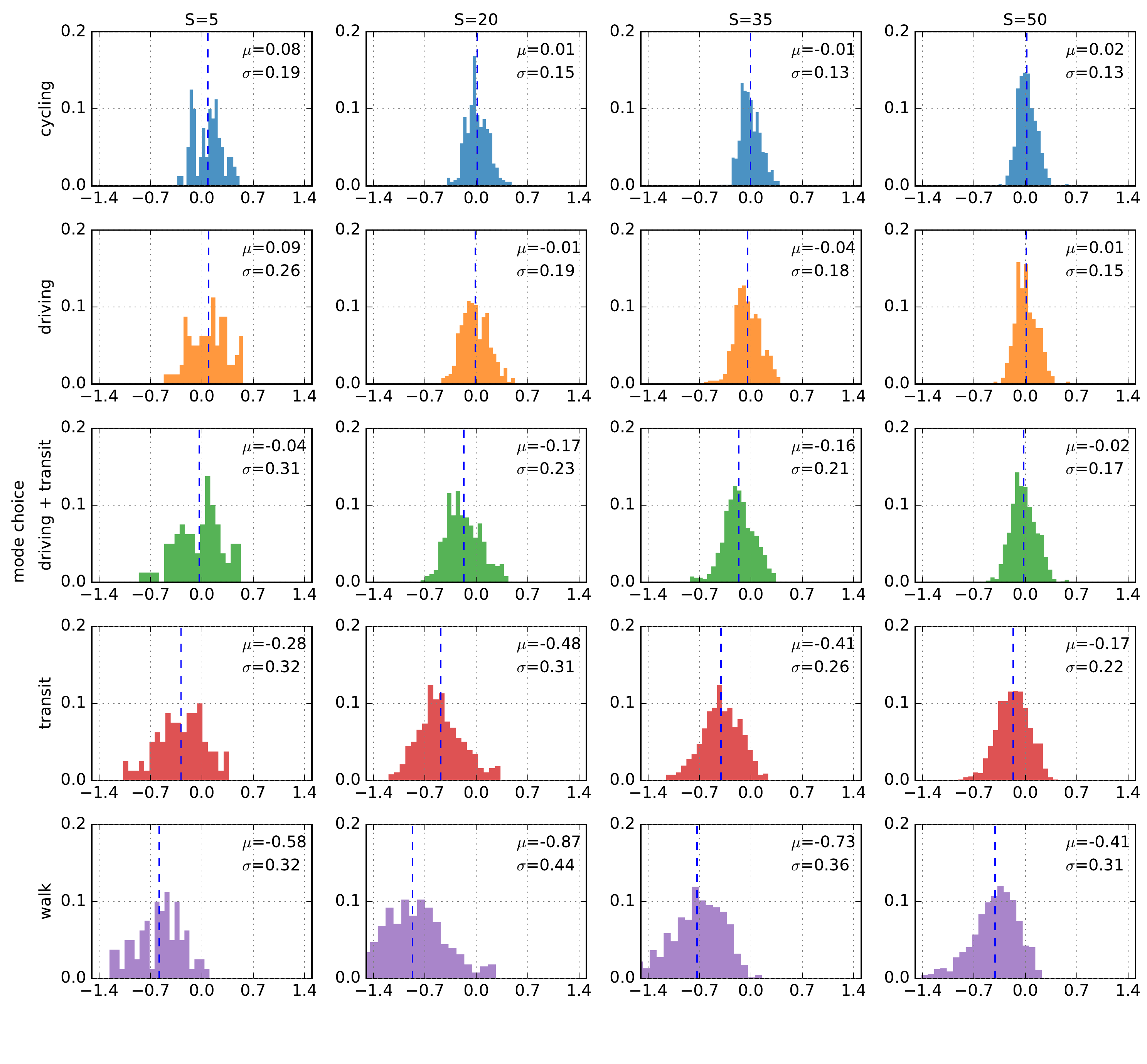}
    \caption{Distribution of data generating parameters. }
    \label{fig:activation}
\end{figure}

The plots in \cref{fig:activation} show the mean and variance of estimated latent variable parameters \(\mathbf{W}'_{hj}\) given the choice outputs.
Since we use binary coding for latent variables, the parameters offer insights into how many latent variables are utilized at any one time.
Parameter vectors with mean values close to zero and low variance indicate that the latent components are sparsely distributed.
We assume that overcomplete representation (\(S\ge X\)) does not cause model overfitting as not all latent variables are active.
The figure shown below illustrates that our generative modelling approach is an efficient method of capturing the underlying heterogeneity across different mode choice decisions.
The mean converges to zero and standard deviation decreases as the number of latent variables increase, indicating that the generative model `suppresses' the influence of less relevant latent variables on the behaviour model.

The results suggest that the RBM learning algorithm inhibits weight connections between the observed and latent variables in order to produce sparse representation.
At (\(S=50\)), the mean parameter activation is near zero with small standard deviation (\(\mu\leq 0.02, \sigma\leq 0.17\)) for \textit{cycling}, \textit{driving}, \textit{driving + transit} modes with an average latent variable activation rate of 85.4\%, 84.4\% and 87.7\% respectively.
For \textit{transit} and \textit{walk} modes the average activation rates are 90.6\% and 92.6\% respectively, indicating that these modes have a higher level of information heterogeneity and less correlated with the observed explanatory variables.

\subsection{Generative model evaluation} 
\begin{figure}[!t]
    \centering
    \includegraphics[width=.75\textwidth]{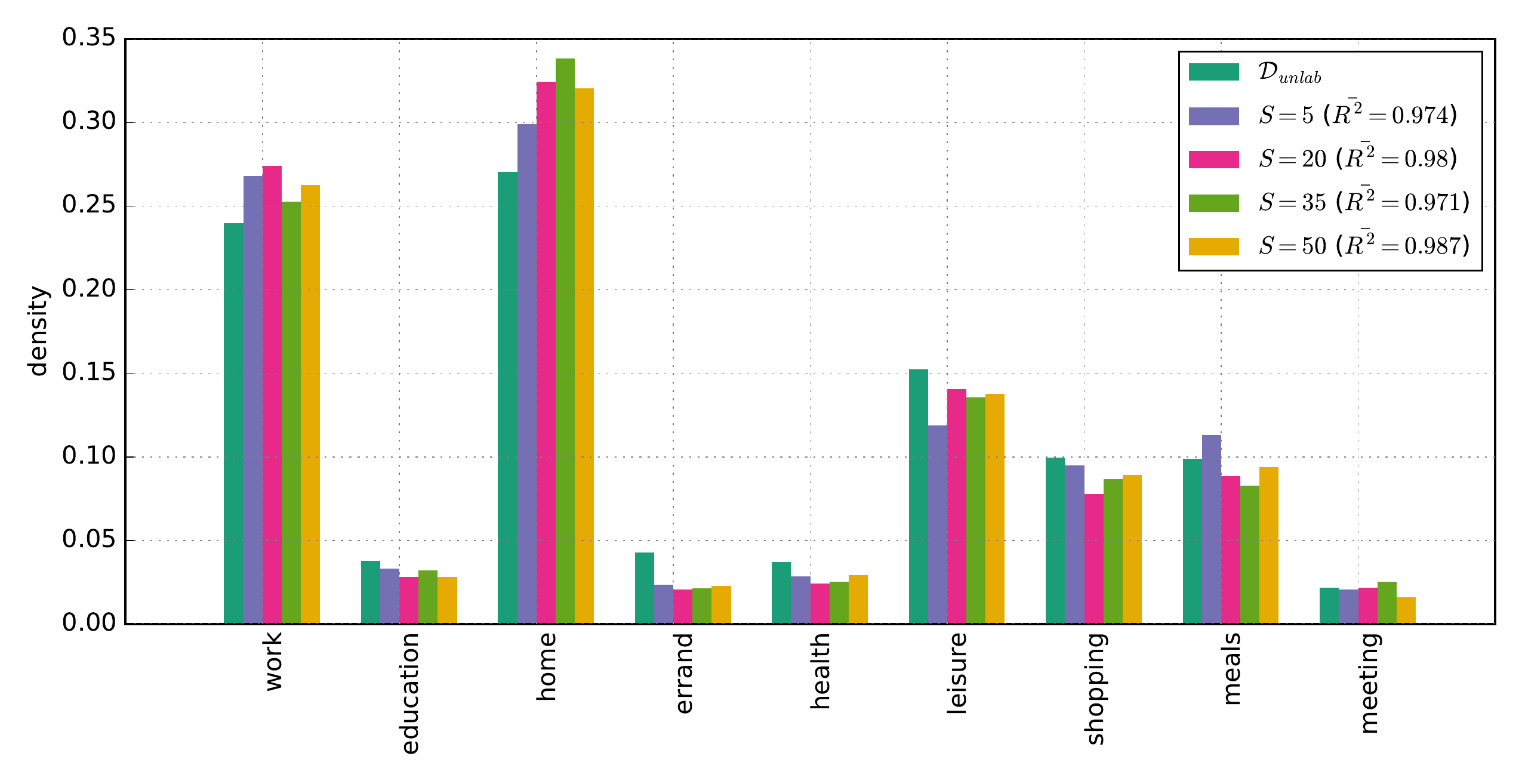}
    \caption{Comparison of data generating output on activity type data.}
    \label{fig:sample_purp}
\end{figure}

\begin{figure}[!t]
    \centering
    \includegraphics[width=.75\textwidth]{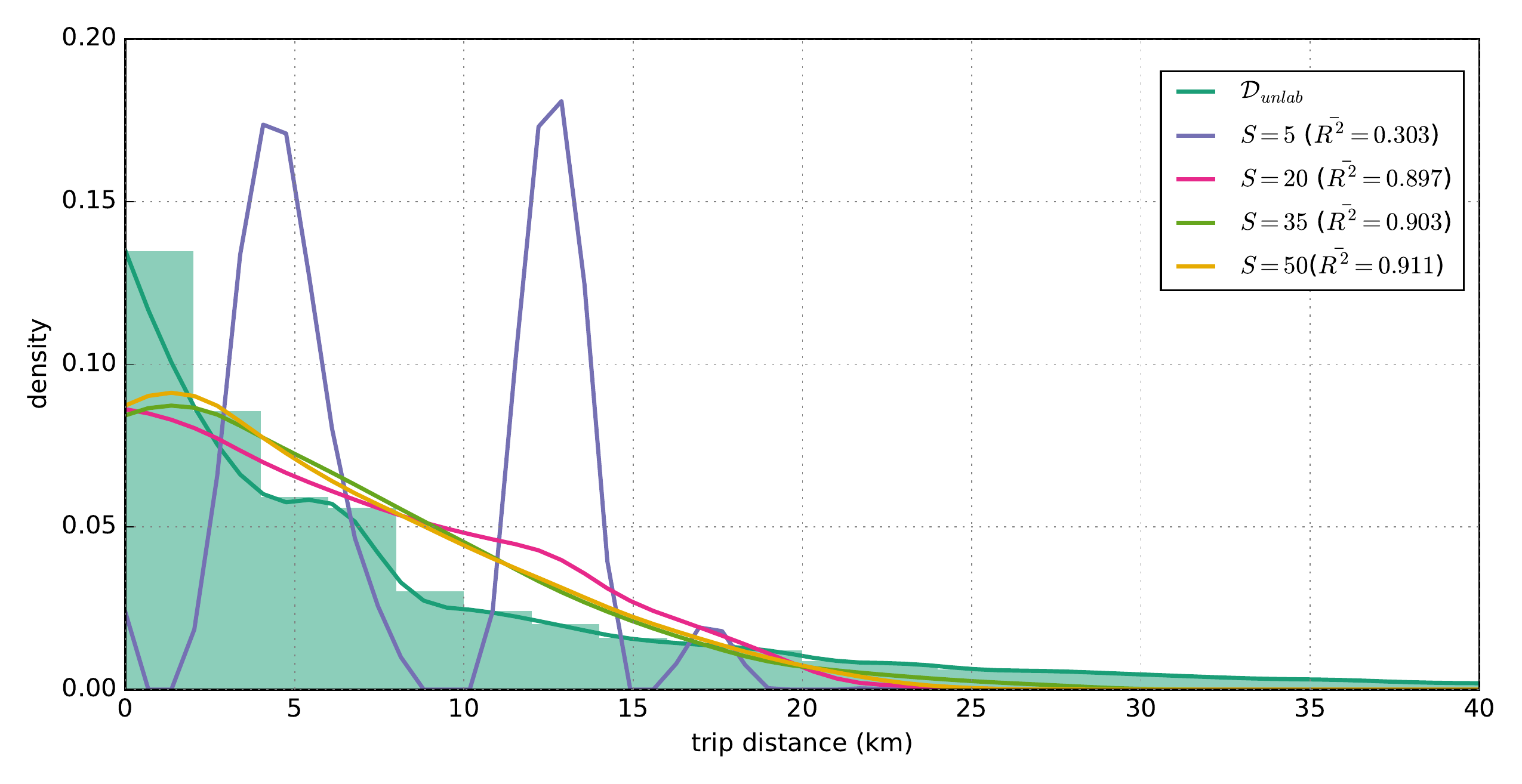}
    \caption{Comparison of data generating output on trip distance data.}
    \label{fig:sample_tripkm}
\end{figure}

\begin{figure}[!t]
    \centering
    \includegraphics[width=.75\textwidth]{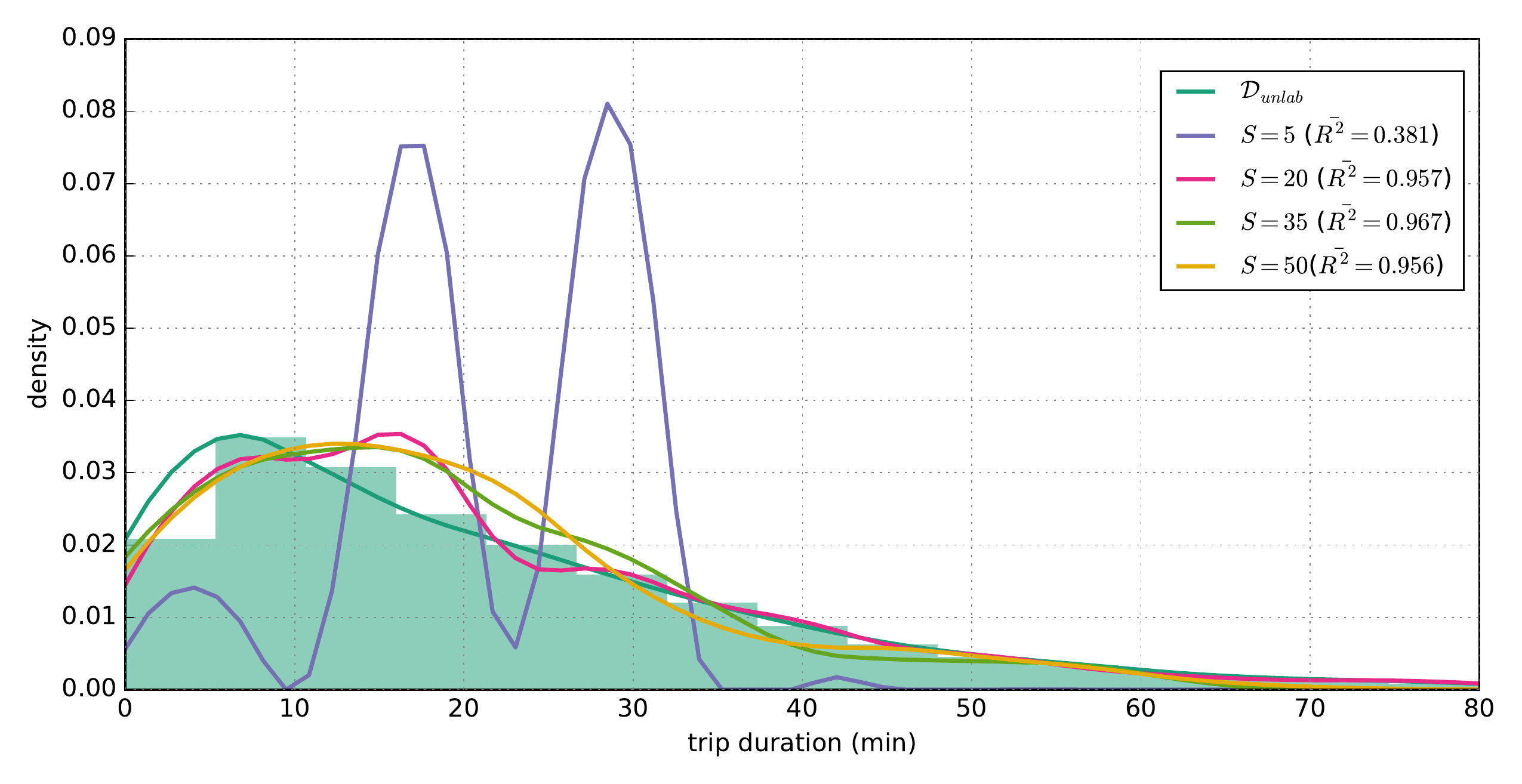}
    \caption{Comparison of data generating output on trip duration data.}
    \label{fig:sample_tripduration}
\end{figure}

To evaluate generative model performance, we measure the statistical fit of the reconstructed distribution.
Simulated reproduction of population data have been used previously to analyze the efficiency of model-based fitting \citep{farooqetal2013}.
Simulation experiments allow evaluation of the model on limited data knowledge, reproducing accurate data distribution while having partial information shows flexibility in capturing decision heterogeneity due to information constraints.
Therefore, the performance results of these simulation experiments can be used to calibrate large scale data-driven models where complex data correlation is present and accounts for the presumption that individuals have limited information processing capacity in choice selection.
We use Gibbs sampling to obtain data from the generative model.
First, evaluate the data generating distribution accuracy using the unlabelled dataset \(\mathcal{D}_{unlab}\).
\cref{fig:sample_purp}, \cref{fig:sample_tripkm} and \cref{fig:sample_tripduration} shows the data generation results for activity, distance and trip duration variables respectively.

Next, for the data generating process, we draw an initial sample from the dataset and fix the observed variable to that data vector and perform Gibbs sampling, alternating between the latent and observed sample conditional probabilities.
Lastly, we clamp the non-target variables to the data vector and update the simulated values of the target observed variable.
For instance, we generate activity type data using the following steps:

\begin{align*}
    \{\tilde{s}_1,...,\tilde{s}_h\} &\sim p(s_1,...,s_h|\textrm{speed,duration,dist,origin,destination}),\\
    \tilde{x} &\sim p(\textrm{activity}|\{\tilde{s}_1,...,\tilde{s}_h\})
\end{align*}

The simulation results show the effects of increasing latent variables on the performance of the data generating model.
\(S=35\) and \(S=50\) achieved high similarities in recovering the original data distribution with \(\bar{R^2}\) value well above \(0.9\).
At \(S=5\), there was an insufficient number of latent variables to capture the structure of the data, shown by the low \(\bar{R^2}\) value. 
Increasing to \(S=20\) significantly improves the result as it increases the non-linear information capacity.

\subsection{Sensitivity analysis of model parameters}
Finally, in this section, we investigate the systematic effects if the generative framework on \(\beta\)-parameters in the mode choice model.
In practice, bias and variances are subject to independent processes, as such, each individual may have vastly different underlying error correction function for the same utility and for each configuration of explanatory variables.
Mixed Logit specification have been used previously to account for this problem, but unfortunately, any variability or noise in the dataset (e.g. through different collection techniques, missing information etc.) will be added to the \(\beta\)-parameter model predictors.
This is less of a problem if one is only interested in the relative variance given the model parameters.
To account for the systematic effects of information heterogeneity, the net utility of each alternative should remain homogeneous across the population (e.g. zero noise level), such that the degree of uncertainty can be compensated by the latent constructs.

\begin{figure}[t]
    \centering
    \includegraphics[width=\textwidth]{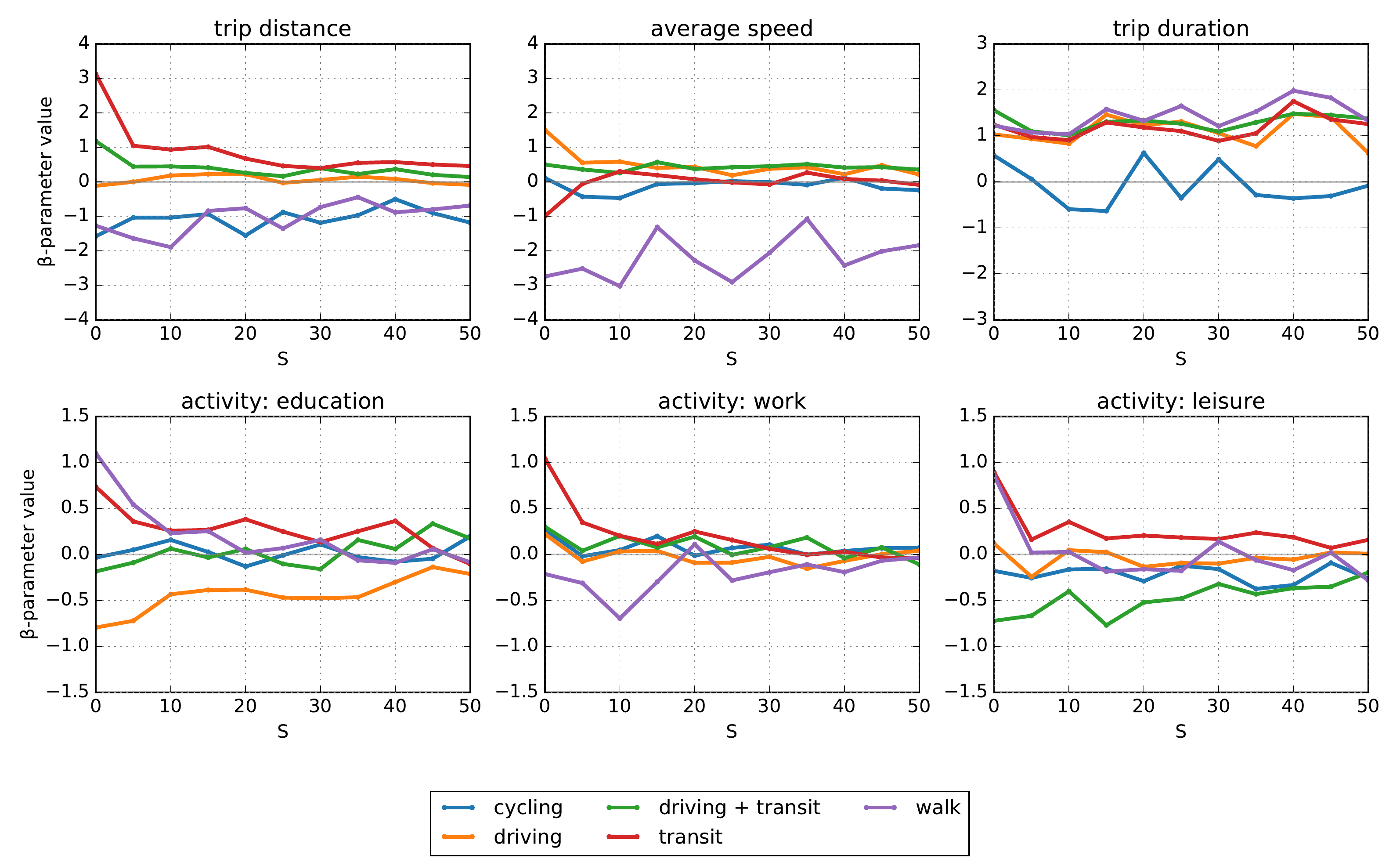}
    \caption{\(\beta\)-parameter estimates using mode choice as the dependent variable, horizontal axis represent number of latent variables.}
    \label{fig:beta}
\end{figure}

\cref{fig:beta} shows the estimated \(\beta\)-parameters of the choice models with different number of latent variables.
The \(\beta\)-parameters identify the systematic effects of each explanatory variable on each choice alternative.
The values on the left edge of each plot show the \(\beta\)-parameters estimated with a standard MNL model.
As we increase the generative model capacity (by increasing the number of latent variables), \(\beta\)-parameters converge to a stable predictor. 
This is an interesting finding as it may in fact indicate that a ordinary utility based choice models may not take into account the systematic effect of information heterogeneity.

We perform a test on the identification of the \(\beta\)-parameters by computing the maximum entropy (maxent) estimate on the observed choice probability in the dataset shown in \cref{tab:ce}. 
The maxent estimate value quantifies the degree of uncertainty within the underlying model accounting for the complexity as well as to determine whether the variance can be attributed to information heterogeneity.
Analysis of maxent can provide information about the uncertainty of the predictors across choice probabilities \cite{golanetal1996}.
We compute maxent of the explanatory variable parameters using the formula:

\begin{equation}
    maxent(\beta_j) = - \sum_j p(y_j)\log \hat{p(y_j)} = - \sum_j p(y_j) \log \Big(\frac{e^{\beta_j}}{\sum_{j'}e^{\beta_{j'}}}\Big)
\end{equation}

where the population class share for each alternatives \(p(y_j)\) are: cycling{=0.068}, driving{=0.613}, driving + transit={0.028}, transit{=0.222} and walking{=0.069} from the labelled dataset.
The resulting \(maxent(\beta_j)\) may therefore be interpreted as the maxent estimate of \(\beta_j\) as the proportion of the sample population in alternative \(j\).
Likewise, a high maxent value indicates a high degree of stochasticity in the decision-making process.
We find \(\hat{p(y_j)}\) by computing \((e^{\beta_j} / \sum_{j'}e^{\beta_{j'}})\).
As the negative entropy increases, e.g. \(maxent(\beta_j) \rightarrow 0\), the correlation between the \(\beta\)-parameter and choice probability converges to the true value, e.g. \(\hat{p(y_j)} \rightarrow p(y_j)\).

The maxent estimate indicates the level of correlation between the set of \(\beta\)-parameters and the output dependent choice variable.
\cref{tab:ce} shows that the \(\beta\)-parameters for distance (2.833) and education activity (2.234) variables in the benchmark model are less likely to influence decisions relative to the other predictors and becomes an indicator of model misspecification.
However, as we increase the number of latent variables in the generative model, maxent decreases and as such, the \(\beta\)-parameters becomes a better predictor of the behaviour.
Evidently, this suggests that the mode choice decision behaviour of individuals are less sensitive trip distance and education related activities.

\begin{table}[!t]
    \centering
    \caption{Result of maxent estimates on \(\beta\)-parameters.}
    \begin{tabu}{X[0.5]X[0.1]X[0.1]X[0.1]X[0.1]X[0.1]}
        \toprule
        Parameters \(\beta_j\)    & \multicolumn{5}{c}{\(maxent(\beta_j\))} \\
        \cmidrule{2-6}
                            & {S=0}  & {S=5} & {S=20} & {S=35} & {S=50} \\
        \midrule
        Distance            & 2.833 & 1.721 & 1.511  & 1.518  & 1.568  \\
        Trip duration       & 1.706 & 1.600 & 1.591  & 1.807  & 1.847  \\
        Speed               & 1.532 & 1.456 & 1.513  & 1.503  & 1.538  \\
        Activity: Edu.      & 2.234 & 2.038 & 1.781  & 1.834  & 1.756  \\
        Activity: Work      & 1.640 & 1.619 & 1.693  & 1.696  & 1.584  \\
        Activity: Leisure   & 1.677 & 1.596 & 1.538  & 1.517  & 1.512  \\
        \midrule
        mean (std. dev.) & 1.94 (0.502) & 1.67 (0.199) & 1.61 (0.11) & 1.65 (0.153) & 1.63 (0.135)\\
        \bottomrule
    \end{tabu}
    \label{tab:ce}
\end{table}

The econometric interpretation of this result implies that individuals seek to use their own prior information (e.g. past experiences, habits, choice dynamics) for mode choice decision rather than driven by exogenous variables.
The significance of the distortion effect of information heterogeneity on the \(\beta\)-parameters decreases as we include a larger correction in the utility function.
This apparent correlation provides evidence that in order to maximize utility and therefore better model prediction accuracy, latent variables can be incorporated in the framework to model information heterogeneity -- The generative model accounts for the variational effects from information heterogeneity, increasing regularity in the utility specification.

Consequently, the estimated \(\beta\)-parameters would reflect the true underlying predictors.
As observed earlier that expected utility can be modelled by the individual's decision strategy shown by evaluating entropy (by a function of latent state vectors) of the choice model.

\section{Discussions and Conclusion}
\label{sec:conclusion}

\subsection{Discussions}
Our findings have several important policy implications. 
First, we have shown that by optimizing a set of internal latent variables to represent distinctive decision strategies of each individual, we can emulate information processing and learning based decision making behaviour incorporated into a choice model. 
We tested the framework and learning algorithm on the dataset to emulate information processing constraints in travel behaviour and decision making.
Our methodology consists of applying an entropy based error component that useed latent constructs in a generative learning model to optimize a set of parameters that minimizes a divergence between the observed and simulated data.

Second, in accordance with behaviour theory in discrete choice analysis, our generative model showed that individuals may not always be utility maximizers and therefore MNL models alone may not be sufficient in modelling travel behaviour in large scale datasets. 
We we have shown that maxent estimates of \(\beta\)-parameters can be reduced by having a learning model component that captures information heterogeneity, population and decision level variance and incorporating the entropy function into choice utilities.
Our analysis and simulating experiments have shown that \(\beta\)-parameter estimates in \cref{fig:beta} scales according to the number of latent variable in the model and it shows significant improvements to choice probability predictions.
The learning framework was able to extract useful information from the dataset, with the assumption that information heterogeneity are present in the data.
The changes in maxent shown in \cref{tab:ce} indicated that the \(\beta\)-parameter have a high level of information heterogeneity, and the misspecification are minimized by incorporating latent variables through a learning process emulated by a generative model.
The explanation for this phenomenon was motivated by information theory: breaking down the processing costs of information related to the choice into a linearly separable component serves as a regularization term in the utility specification.

Lastly, it would suggest that distance based trip planning are more strongly correlated to long term individual habits and perception of the travel route and less likely due to explicit change in trip distance. 
Our experiments showed how some explanatory variables can contain a larger source of information heterogeneity and increasing the generative model capacity increases the choice probability accuracy more robustly. 
The results indicated improved model fit can be attributed to more efficient use of the generative model, which suggests that stochastic choice selection in decision making can be associated to the availability individual's prior information.

\subsection{Conclusion}
Generative modelling presents a new perspective on how analysts can obtain insights into behavioural heterogeneity manifestations by accounting for information processing constraints in the model learning process.
Based on rational inattention behaviour and information theory, we develop a systematic approach to identify information heterogeneity and we propose a data-driven generative learning process to emulate decision making under uncertainty and information processing constraints.
It explains why not all exogenous information are used in the decision making process as discussed in \citep{sims2010}.

The impact of this study on travel demand modelling is that we can take advantage of noisy data (e.g. GPS, Wi-Fi, cellular networks) to develop a flexible, operational, and adaptive model framework.
Our basic assumption is that large and unstructured data from passive information sources which contain behavioural information not captured in explanatory variables, can be exploited with the proper learning models and optimization algorithms.
This study demonstrates the properties and descriptive power of the generative modelling framework to emulate decisions under uncertainty and information processing constraints.
We define the source of heterogeneity to be the inherent nature of the data itself, and by updating the model using an iterative KL divergence minimization process we are able to synthetically reproduce the unobserved variations using latent constructs in a generative model.
The latent constructs provide additional error correction for information heterogeneity in the utility specification, allowing the model to simulate decision making and choice actions with internal information processing components.
It also allows a convenient representation of entropy, by incorporating an error generating function into the framework.
Our results indicate a strong correlation with rational inattention behaviour theory, which shows that individuals may tend to ignore certain explanatory variables or rely on prior information for discrete choice decision making.
The experiments identify several important components of the generative model which are more sensitive to information heterogeneity and applies an automatic correction for this variation by representing the heterogeneity as an entropy measure in the utility specification.
More generally, principles from generative modelling demonstrated in this paper can be applied to existing travel behaviour analysis to benefit from using large data sources, where latent behaviour information are not directly captured in the explanatory variables.

\subsection{Future work}
The scope of this paper focuses on the implementation and basic methodology of developing a machine learning based generative model for discrete choice analysis.
There are several extensions to this study which can be addressed in future work:
\begin{enumerate}
    \item[(i)]  Exploring the use of variation inference techniques in Mixed Logit models to address estimation tractability, allowing for a comparative analysis between discrete choice and machine learning based methods.
    \item[(ii)]Several other variants of generative model learning algorithms (e.g. GANs, Autoencoders) could be tested to gain insights into how they would emulate different social and cognitive behavioural concepts.
    Additionally, generative modelling can be expanded to other constraints beyond information processing costs, for example, budget and time constraints.
\end{enumerate}

\section*{Acknowledgements}
This research is funded by Canada Research Chair in Disruptive Transportation Technologies and Services and Ryerson University.

\section*{References}
\bibliographystyle{elsarticle-num}
\bibliography{bibliography.bib}

\begin{thebibliography}{10}
\expandafter\ifx\csname url\endcsname\relax
  \def\url#1{\texttt{#1}}\fi
\expandafter\ifx\csname urlprefix\endcsname\relax\def\urlprefix{URL }\fi
\expandafter\ifx\csname href\endcsname\relax
  \def\href#1#2{#2} \def\path#1{#1}\fi

\bibitem{mcfaddentrain2000}
D.~McFadden, K.~Train, Mixed {MNL} models for discrete response, Journal of
  applied Econometrics 15~(5) (2000) 447--470.
\newblock \href
  {http://dx.doi.org/10.1002/1099-1255(200009/10)15:5<447::AID-JAE570>3.0.CO;2-1}
  {\path{doi:10.1002/1099-1255(200009/10)15:5<447::AID-JAE570>3.0.CO;2-1}}.

\bibitem{lietal2016}
D.~Li, T.~Miwa, T.~Morikawa, P.~Liu, Incorporating observed and unobserved
  heterogeneity in route choice analysis with sampled choice sets,
  Transportation Research Part C: Emerging Technologies 67 (2016) 31--46.
\newblock \href {http://dx.doi.org/10.1016/j.trc.2016.02.002}
  {\path{doi:10.1016/j.trc.2016.02.002}}.

\bibitem{vijkrueger2017}
A.~Vij, R.~Krueger, Random taste heterogeneity in discrete choice models:
  Flexible nonparametric finite mixture distributions, Transportation Research
  Part B: Methodological 106 (2017) 76--101.
\newblock \href {http://dx.doi.org/10.1016/j.trb.2017.10.013}
  {\path{doi:10.1016/j.trb.2017.10.013}}.

\bibitem{nikolicbierlaire2017}
M.~Nikoli{\'c}, M.~Bierlaire, Data-driven spatio-temporal discretization for
  pedestrian flow characterization, Transportation research procedia 23 (2017)
  188--207.
\newblock \href {http://dx.doi.org/10.1016/j.trc.2017.08.026}
  {\path{doi:10.1016/j.trc.2017.08.026}}.

\bibitem{gopinath1994}
D.~A. Gopinath, Modeling heterogeneity in discrete choice processes:
  Application to travel demand, Ph.D. thesis, MIT (1995).

\bibitem{bolducalvarezdaziano2010}
D.~Bolduc, R.~Alvarez-Daziano, On estimation of hybrid choice models, in:
  S.~Hess, A.~Daly (Eds.), Choice Modelling: The State-of-the-art and The
  State-of-practice, Edward Elgar, 2010, pp. 259--287.
\newblock \href {http://dx.doi.org/10.1108/9781849507738-011}
  {\path{doi:10.1108/9781849507738-011}}.

\bibitem{wong2018modelling}
M.~Wong, B.~Farooq, Modelling latent travel behaviour characteristics with
  generative machine learning, in: 2018 21st International Conference on
  Intelligent Transportation Systems (ITSC), 2018, pp. 749--754.
\newblock \href {http://dx.doi.org/10.1109/ITSC.2018.8569581}
  {\path{doi:10.1109/ITSC.2018.8569581}}.

\bibitem{wong2018discriminative}
M.~Wong, B.~Farooq, G.-A. Bilodeau, Discriminative conditional restricted
  boltzmann machine for discrete choice and latent variable modelling, Journal
  of choice modelling 29 (2018) 152--168.
\newblock \href {http://dx.doi.org/10.1016/j.jocm.2017.11.003}
  {\path{doi:10.1016/j.jocm.2017.11.003}}.

\bibitem{cherchipolak2005}
E.~Cherchi, J.~W. Polak, Assessing user benefits with discrete choice models:
  Implications of specification errors under random taste heterogeneity,
  Transportation Research Record 1926~(1) (2005) 61--69.
\newblock \href {http://dx.doi.org/10.1177/0361198105192600108}
  {\path{doi:10.1177/0361198105192600108}}.

\bibitem{sims2010}
C.~A. Sims, Rational inattention and monetary economics, in: B.~M. Friedman,
  M.~Woodford (Eds.), Handbook of monetary economics, Vol.~3, Elsevier, 2010,
  Ch.~4, pp. 155--181.
\newblock \href {http://dx.doi.org/10.1016/B978-0-444-53238-1.00004-1}
  {\path{doi:10.1016/B978-0-444-53238-1.00004-1}}.

\bibitem{matvejkamckay2015}
F.~Mat{\v{e}}jka, A.~McKay, Rational inattention to discrete choices: A new
  foundation for the multinomial logit model, American Economic Review 105~(1)
  (2015) 272--98.
\newblock \href {http://dx.doi.org/10.1257/aer.20130047}
  {\path{doi:10.1257/aer.20130047}}.

\bibitem{alizadeh2018online}
H.~Alizadeh, P.-L. Bourbonnais, C.~Morency, B.~Farooq, N.~Saunier, An online
  survey to enhance the understanding of car drivers route choices,
  Transportation Research Procedia 32 (2018) 482--494.
\newblock \href {http://dx.doi.org/10.1016/j.trpro.2018.10.042}
  {\path{doi:10.1016/j.trpro.2018.10.042}}.

\bibitem{fosgerauetal2017}
M.~Fosgerau, E.~Melo, A.~d. Palma, M.~Shum, Discrete choice and rational
  inattention: A general equivalence result, arXiv preprint arXiv:1709.09117.

\bibitem{fosgeraujiang2019}
M.~Fosgerau, G.~Jiang, Travel time variability and rational inattention,
  Transportation Research Part B: Methodological 120 (2019) 1--14.
\newblock \href {http://dx.doi.org/10.1016/j.trb.2018.12.003}
  {\path{doi:10.1016/j.trb.2018.12.003}}.

\bibitem{sims2003}
C.~A. Sims, Implications of rational inattention, Journal of monetary Economics
  50~(3) (2003) 665--690.
\newblock \href {http://dx.doi.org/10.1016/S0304-3932(03)00029-1}
  {\path{doi:10.1016/S0304-3932(03)00029-1}}.

\bibitem{friston2006}
K.~Friston, J.~Kilner, L.~Harrison, A free energy principle for the brain,
  Journal of Physiology-Paris 100~(1-3) (2006) 70--87.
\newblock \href {http://dx.doi.org/10.1016/j.jphysparis.2006.10.001}
  {\path{doi:10.1016/j.jphysparis.2006.10.001}}.

\bibitem{ellsberg1961}
D.~Ellsberg, Risk, ambiguity, and the savage axioms, The quarterly journal of
  economics (1961) 643--669\href {http://dx.doi.org/10.2307/1884324}
  {\path{doi:10.2307/1884324}}.

\bibitem{kahnemantversky1979}
D.~Kahneman, A.~Tversky, Prospect theory: An analysis of decision under risk,
  Econometrica 47~(2) (1979) 263--292.
\newblock \href {http://dx.doi.org/10.1142/9789814417358_0006}
  {\path{doi:10.1142/9789814417358_0006}}.

\bibitem{steineretal2017}
J.~Steiner, C.~Stewart, F.~Mat{\v{e}}jka, Rational inattention dynamics:
  Inertia and delay in decision-making, Econometrica 85~(2) (2017) 521--553.
\newblock \href {http://dx.doi.org/10.3982/ECTA13636}
  {\path{doi:10.3982/ECTA13636}}.

\bibitem{teyeetal2017}
C.~Teye, M.~G. Bell, M.~C. Bliemer, Entropy maximising facility location model
  for port city intermodal terminals, Transportation Research Part E: Logistics
  and Transportation Review 100 (2017) 1--16.
\newblock \href {http://dx.doi.org/10.1016/j.tre.2017.01.006}
  {\path{doi:10.1016/j.tre.2017.01.006}}.

\bibitem{leard2018}
B.~Leard, Consumer inattention and the demand for vehicle fuel cost savings,
  Journal of choice modelling 29 (2018) 1--16.
\newblock \href {http://dx.doi.org/10.1016/j.jocm.2018.08.002}
  {\path{doi:10.1016/j.jocm.2018.08.002}}.

\bibitem{anas1983}
A.~Anas, Discrete choice theory, information theory and the multinomial logit
  and gravity models, Transportation Research Part B: Methodological 17~(1)
  (1983) 13--23.
\newblock \href {http://dx.doi.org/10.1016/0191-2615(83)90023-1}
  {\path{doi:10.1016/0191-2615(83)90023-1}}.

\bibitem{ullah1996}
A.~Ullah, Entropy, divergence and distance measures with econometric
  applications, Journal of Statistical Planning and Inference 49~(1) (1996)
  137--162.
\newblock \href {http://dx.doi.org/10.1016/0378-3758(95)00034-8}
  {\path{doi:10.1016/0378-3758(95)00034-8}}.

\bibitem{goodfellowetal2016}
I.~Goodfellow, Y.~Bengio, A.~Courville, Deep Learning, MIT Press, 2016,
  \url{http://www.deeplearningbook.org}.

\bibitem{hintonsalakhutdinov2006}
G.~E. Hinton, R.~R. Salakhutdinov, Reducing the dimensionality of data with
  neural networks, science 313~(5786) (2006) 504--507.

\bibitem{hintonetal2006}
G.~E. Hinton, S.~Osindero, Y.-W. Teh, A fast learning algorithm for deep belief
  nets, Neural computation 18~(7) (2006) 1527--1554.
\newblock \href {http://dx.doi.org/10.1162/neco.2006.18.7.1527}
  {\path{doi:10.1162/neco.2006.18.7.1527}}.

\bibitem{wongfarooq2019}
M.~Wong, B.~Farooq, A bi-partite generative model framework for analyzing and
  simulating large scale multiple discrete-continuous travel behaviour data,
  arXiv preprint arXiv:1901.06415.

\bibitem{ranzatoetal2007}
M.~Ranzato, C.~Poultney, S.~Chopra, Y.~LeCun, Efficient learning of sparse
  representations with an energy-based model, in: Advances in neural
  information processing systems, 2007, pp. 1137--1144.

\bibitem{kingmawelling2013}
D.~P. Kingma, M.~Welling, Auto-encoding variational bayes, arXiv preprint
  arXiv:1312.6114.

\bibitem{bleietal2017}
D.~M. Blei, A.~Kucukelbir, J.~D. McAuliffe, Variational inference: A review for
  statisticians, Journal of the American Statistical Association 112~(518)
  (2017) 859--877.
\newblock \href {http://dx.doi.org/10.1080/01621459.2017.1285773}
  {\path{doi:10.1080/01621459.2017.1285773}}.

\bibitem{train2009}
K.~E. Train, Discrete choice methods with simulation, Cambridge university
  press, 2009.
\newblock \href {http://dx.doi.org/10.1017/CBO9780511805271}
  {\path{doi:10.1017/CBO9780511805271}}.

\bibitem{alwosheeletal2018}
A.~Alwosheel, S.~van Cranenburgh, C.~G. Chorus, Is your dataset big enough?
  sample size requirements when using artificial neural networks for discrete
  choice analysis, Journal of choice modelling 28 (2018) 167--182.
\newblock \href {http://dx.doi.org/10.1016/j.jocm.2018.07.002}
  {\path{doi:10.1016/j.jocm.2018.07.002}}.

\bibitem{heetal2016}
K.~He, X.~Zhang, S.~Ren, J.~Sun, Deep residual learning for image recognition,
  in: Proceedings of the IEEE conference on computer vision and pattern
  recognition, 2016, pp. 770--778.

\bibitem{tehetal2003}
Y.~W. Teh, M.~Welling, S.~Osindero, G.~E. Hinton, Energy-based models for
  sparse overcomplete representations, Journal of Machine Learning Research 4
  (2003) 1235--1260.

\bibitem{datamobile2016}
{Ville de montr\'{e}al},
  \href{https://ville.montreal.qc.ca/mtltrajet/}{D\'{e}placements {MTL} trajet}
  (2016).
\newline\urlprefix\url{https://ville.montreal.qc.ca/mtltrajet/}

\bibitem{ranzatoetal2008}
M.~Ranzato, Y.-l. Boureau, Y.~LeCun, Sparse feature learning for deep belief
  networks, in: Advances in neural information processing systems, 2008, pp.
  1185--1192.

\bibitem{glorotetal2011}
X.~Glorot, A.~Bordes, Y.~Bengio, Deep sparse rectifier neural networks, in:
  Proceedings of the fourteenth international conference on artificial
  intelligence and statistics, 2011, pp. 315--323.

\bibitem{farooqetal2013}
B.~Farooq, M.~Bierlaire, R.~Hurtubia, G.~Fl{\"o}tter{\"o}d, Simulation based
  population synthesis, Transportation Research Part B: Methodological 58
  (2013) 243--263.
\newblock \href {http://dx.doi.org/10.1016/j.trb.2013.09.012}
  {\path{doi:10.1016/j.trb.2013.09.012}}.

\bibitem{golanetal1996}
A.~Golan, G.~Judge, J.~M. Perloff, A maximum entropy approach to recovering
  information from multinomial response data, Journal of the American
  Statistical Association 91~(434) (1996) 841--853.

\end{thebibliography}

\end{document}